\documentclass[useAMS,usenatbib]{mn2e}
\usepackage{savesym}
\usepackage{txfonts}
\savesymbol{iint}
\savesymbol{iiint}
\savesymbol{iiiint}
\savesymbol{idotsint}
\usepackage{graphicx}
\usepackage{amssymb}
\usepackage[below]{placeins}
\usepackage{txfonts}
\usepackage{natbib}
\usepackage{amsmath}
\usepackage{float}
\restoresymbol{TXF}{iint}
\restoresymbol{TXF}{iiint}
\restoresymbol{TXF}{iiiint}
\restoresymbol{TXF}{idotsint}
\usepackage{multirow}
\usepackage{array}
\usepackage{xcolor}
\usepackage[percent]{overpic}
\bibpunct{(}{)}{;}{a}{}{,}

\def \chandra{{\emph{Chandra}}}

\def\spose#1{\hbox to 0pt{#1\hss}}
\def\approxlt{\mathrel{\spose{\lower 3pt\hbox{$\sim$}}
        \raise 2.0pt\hbox{$<$}}}
\def\approxgt{\mathrel{\spose{\lower 3pt\hbox{$\sim$}}
        \raise 2.0pt\hbox{$>$}}}
\def\approxpropto{\mathrel{\spose{\lower 3pt\hbox{$\sim$}}
        \raise 2.0pt\hbox{$\propto$}}}
\mathchardef\twiddle="2218

\def\multleft#1{\hbox to size{\vbox {\halign {\lft{##}\cr #1}}\hfill}\par}
\def\multright#1{\hbox to size{\vbox {\halign {\rt{##}\cr #1}}\hfill}\par}

\def\today{\ifcase\month\or January\or February\or March\or April\or May\or
      June\or July\or August\or September\or October\or November\or December\fi
      \space\number\day, \number\year}
\def\<{\thinspace}

\def\arcsec{{\rm\thinspace arcsec}}

\def\chandra{{\it Chandra}}

\def\arcsec {\hbox{$^{\prime\prime}$}}

\makeatletter
\newcommand{\thickhline}{%
    \noalign {\ifnum 0=`}\fi \hrule height 1.2pt
    \futurelet \reserved@a \@xhline
}
\newcolumntype{"}{@{\hskip\tabcolsep\vrule width 1pt\hskip\tabcolsep}}
\makeatother

\newcommand{\ion}[2]{#1\,{\sc{#2}}}

\bibpunct{(}{)}{;}{a}{}{,}

\title[{\it Chandra} study of the Ophiuchus cluster]{Deep {\it Chandra} study of the truncated cool core of the Ophiuchus cluster}

\author[Werner et al.]{N. Werner$^{1,2}$, I. Zhuravleva$^{1,2}$, R. E. A. Canning$^{1,2}$, S. W. Allen$^{1,2,3}$, A. L. King$^{1,2}$, \newauthor J. S. Sanders$^{4}$, A. Simionescu$^{5}$, G. B. Taylor$^{6}$, R. G. Morris$^{1,3}$, A. C. Fabian$^{7}$ \\
$^1$Kavli Institute for Particle Astrophysics and Cosmology, Stanford University, 452 Lomita Mall, Stanford, CA 94305-4085, USA \\
$^2$Department of Physics, Stanford University, 382 Via Pueblo Mall, Stanford, CA 94305-4060, USA \\
$^3$SLAC National Accelerator Laboratory, 2575 Sand Hill Road, Menlo Park, CA 94025, USA\\
$^4$Max-Planck-Institut f\"ur extraterrestrische Physik, Giessenbachstrasse 1, 85748 Garching, Germany\\
$^5$Institute of Space and Astronautical Science (ISAS), JAXA, 3-1-1 Yoshinodai, Chuo-ku, Sagamihara, Kanagawa, 252-5210, Japan\\
$^6$Department of Physics and Astronomy, University of New Mexico, Albuquerque, NM 87131, USA\\
$^7$Institute of Astronomy, Madingley Road, Cambridge CB3 0HA, UK\\
}

\begin{document}
\maketitle

\begin{abstract}
We present the results of a deep (280 ks) {\it Chandra} observation of the Ophiuchus cluster, the second-brightest galaxy cluster in the X-ray sky. 
The cluster hosts a truncated cool core, with a temperature increasing from $kT\sim1$~keV in the core to $kT\sim9$~keV at $r\sim30$~kpc. Beyond $r\sim30$~kpc the intra-cluster medium (ICM) appears remarkably isothermal. The core is dynamically disturbed with multiple sloshing induced cold fronts, with indications for both Rayleigh-Taylor and Kelvin-Helmholtz instabilities. The sloshing is the result of the strongly perturbed gravitational potential in the cluster core, with the central brightest cluster galaxy (BCG) being displaced southward from the global center of mass. The residual image reveals a likely subcluster south of the core at the projected distance of $r\sim280$~kpc. The cluster also harbors a likely radio phoenix, a source revived by adiabatic compression by gas motions in the ICM. Even though the Ophiuchus cluster is strongly dynamically active, the amplitude of density fluctuations outside of the cooling core is low, indicating velocities smaller than $\sim100$ km~s$^{-1}$. The density fluctuations might be damped by thermal conduction in the hot and remarkably isothermal ICM, resulting in our underestimate of gas velocities. We find a surprising, sharp surface brightness discontinuity, that is curved away from the core, at $r\sim120$~kpc to the southeast of the cluster center. We conclude that this feature is most likely due to gas dynamics associated with a merger and not a result of an extraordinary active galactic nucleus (AGN) outburst. The cooling core lacks any observable X-ray cavities and the AGN only displays weak, point-like radio emission, lacking lobes or jets, indicating that currently it may be largely dormant. The lack of strong AGN activity may be due to the bulk of the cooling taking place offset from the central supermassive black hole.
\end{abstract}


\section{Introduction}

The degree to which cool cores are disrupted by merger events is an outstanding question in cluster physics, with implications for galaxy cluster evolution and cosmology \citep[e.g.][]{burns2008,allen2011}. Some simulations have suggested that cluster cool cores are very difficult to destroy once formed, and are therefore expected to survive major mergers \citep{burns2008,poole2008}. However, these conclusions are sensitive to the hydrodynamical/magnetohydrodynamical treatments employed. Our previous analysis of a short 50~ks {\it Chandra} observation of the nearby Ophiuchus cluster also challenges this view \citep{million2010}.

The Ophiuchus cluster \citep[$z=0.0296$,][]{durret2015} is the second-brightest galaxy cluster in the 2--10~keV sky \citep{edge1990}. It is the most massive structure in the Ophiuchus Supercluster and is surrounded by many in-falling smaller clusters and groups \citep{wakamatsu2005,hasegawa2000}. Spatially resolved spectroscopy using the previous {\it Chandra} observation \citep{million2010} revealed remarkably steep temperature and iron abundance gradients within the cool core: in the central 30~kpc, the temperature plunges by almost an order of magnitude, and the metallicity varies by a similar amount. Beyond this radius, however, and out to the edge of the {\it Chandra} field of view, near constant temperature and metallicity are observed. The cool core of the cluster appears to have been truncated. 

The previous {\it Chandra} observations also revealed a series of cold fronts \citep{ascasibar2006,million2010}, suggesting substantial `sloshing' of the X-ray emitting gas \citep{markevitch2001,tittley2005,ascasibar2006,markevitch2007}.  The X-ray emission associated with the cool core is very sharply peaked and the X-ray morphology of the inner 10 kpc region is complex. The X-ray and stellar optical/near-IR brightness peaks are offset by $\sim4$ arcsec \citep[$\sim2.2$ kpc;][]{hamer2012}. These observations indicate that the cool core of the Ophiuchus cluster has been stripped by the violent gas sloshing in the center of the cluster.

Here, we present a deep (280~ks) {\it Chandra} observation of the Ophiuchus cluster along with new high resolution Jansky Very Large Array (JVLA) 1--2 GHz radio data, to study the effects of strong gas dynamics on the evolution of the cool core in detail. At a redshift of $z=0.0296$ \citep{durret2015}, assuming the concordance $\Lambda$CDM cosmology, one arcsecond corresponds to 0.59~kpc. 

\section{Observations and data analysis}
\label{analysis}

\begin{table}
\begin{center}
\caption{Summary of the \chandra\ observations.
Columns list the observation ID, observation date, 
and the exposure after cleaning.}\label{table:obs}
\begin{tabular}{ccc}
\hline\hline
Obs. ID & Obs. date   & Exposure (ks)\\
\hline 
3200& 2002 Oct. 21  & 50.5  \\
16142& 2014 Jul. 1  & 32.6  \\
16143& 2014 Jul. 13   & 14.9 \\
16464& 2014 Aug. 8  & 31.7  \\
16626& 2014 Jul. 3  & 30.8  \\
16627& 2014 Jul. 5  & 34.6  \\
16633& 2014 Aug. 4  & 9.5  \\
16634& 2014 Jul. 11  & 22.4  \\
16635& 2014 Jul. 12   & 18.9  \\
16645& 2014 Aug. 9   & 33.6  \\
\hline
\end{tabular}
\end{center}
\end{table}

\subsection{Chandra data}

The \chandra\ X-ray observatory pointed at the Ophiuchus cluster in July and August  2014 for 230~ks using the Advanced CCD Imaging Spectrometer (ACIS) in the VFAINT mode, using CCDs 1--4 and 6. We also included in our analysis a 50~ks observation performed in October 2002 using the ACIS-S3 chip (CCD-7). 
The standard level-1 event lists were reprocessed using the {\texttt{CIAO}} (version~4.7, {\texttt{CALDB~4.6.5}}) software package, including the latest gain maps and calibration products. Bad pixels were removed and standard grade selections applied. The data were cleaned to remove periods of anomalously high background. The observations, along with the identifiers, dates, and net exposure times after cleaning (total net exposure of 280~ks), are listed in Table 1. Background images and spectra were extracted from the blank-sky fields available from the Chandra X-ray Center. These were cleaned in an identical way to the source observations, reprojected to the same coordinate system and normalized by the ratio of the observed to blank-sky count rates in the 9.5--12~keV band. 

\label{analysis}
\begin{figure*}
\begin{minipage}{0.45\textwidth}
\includegraphics[width=1.17\textwidth,clip=t,angle=0.]{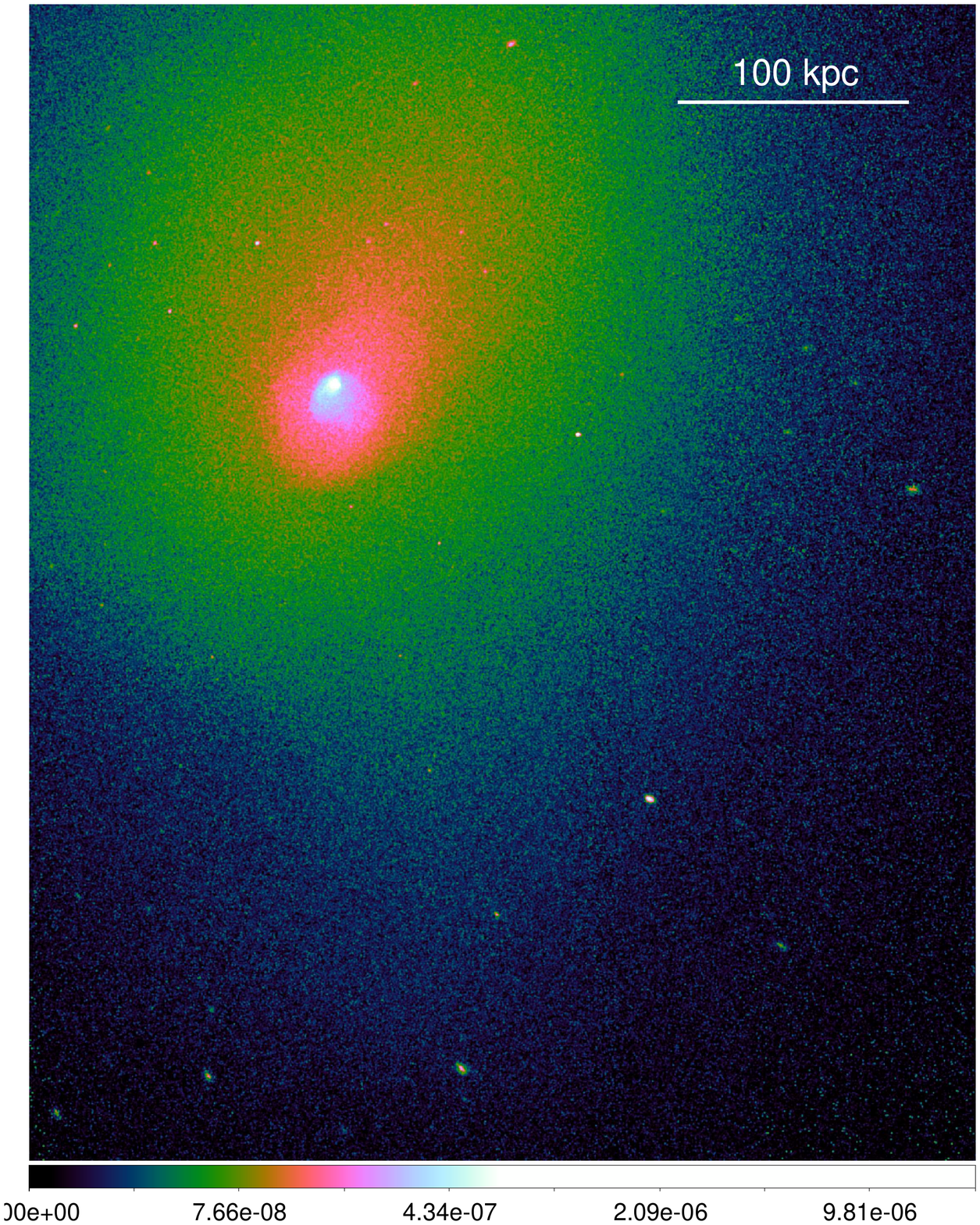}
\end{minipage}
\begin{minipage}{0.45\textwidth}
\hspace{0.7cm}\vspace{-0.45cm}\includegraphics[width=1.035\textwidth,clip=t,angle=0.]{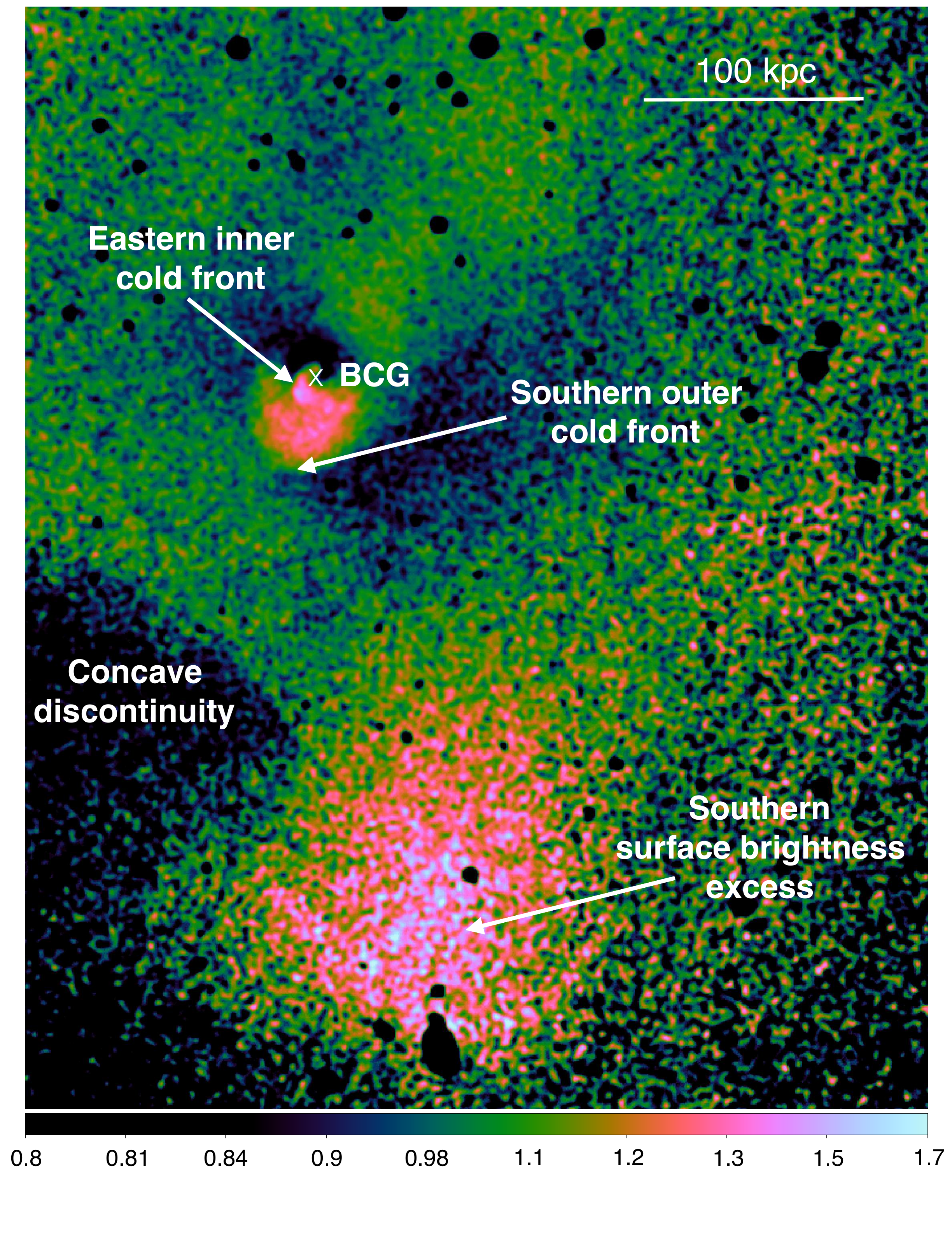}
\end{minipage}
\vspace{-0.7cm}
\caption{{\it Left panel:} \chandra\ X-ray image of the Ophiuchus cluster in the 0.6--7.5~keV energy band. The image was smoothed with a Gaussian function with a 1.0 arcsec window. The colour bar shows the surface brightness, in counts~s$^{-1}$ per $0.492\times0.492$ arcsec pixel.  {\it Right panel:} The matching residual image, obtained by dividing the image on the left with an elliptical double beta-model, reveals a prominent surface brightness excess in the south and a surprising, sharp discontinuity southeast of the cluster center, curved away from the core. North is up and east is left in all images. } 
\label{large_IM}
\end{figure*}

\label{analysis}
\begin{figure*}
\begin{minipage}{0.45\textwidth}
\vspace{-1.2cm}
\hspace{-0.5cm}
\includegraphics[width=1.055\textwidth,clip=t,angle=0.]{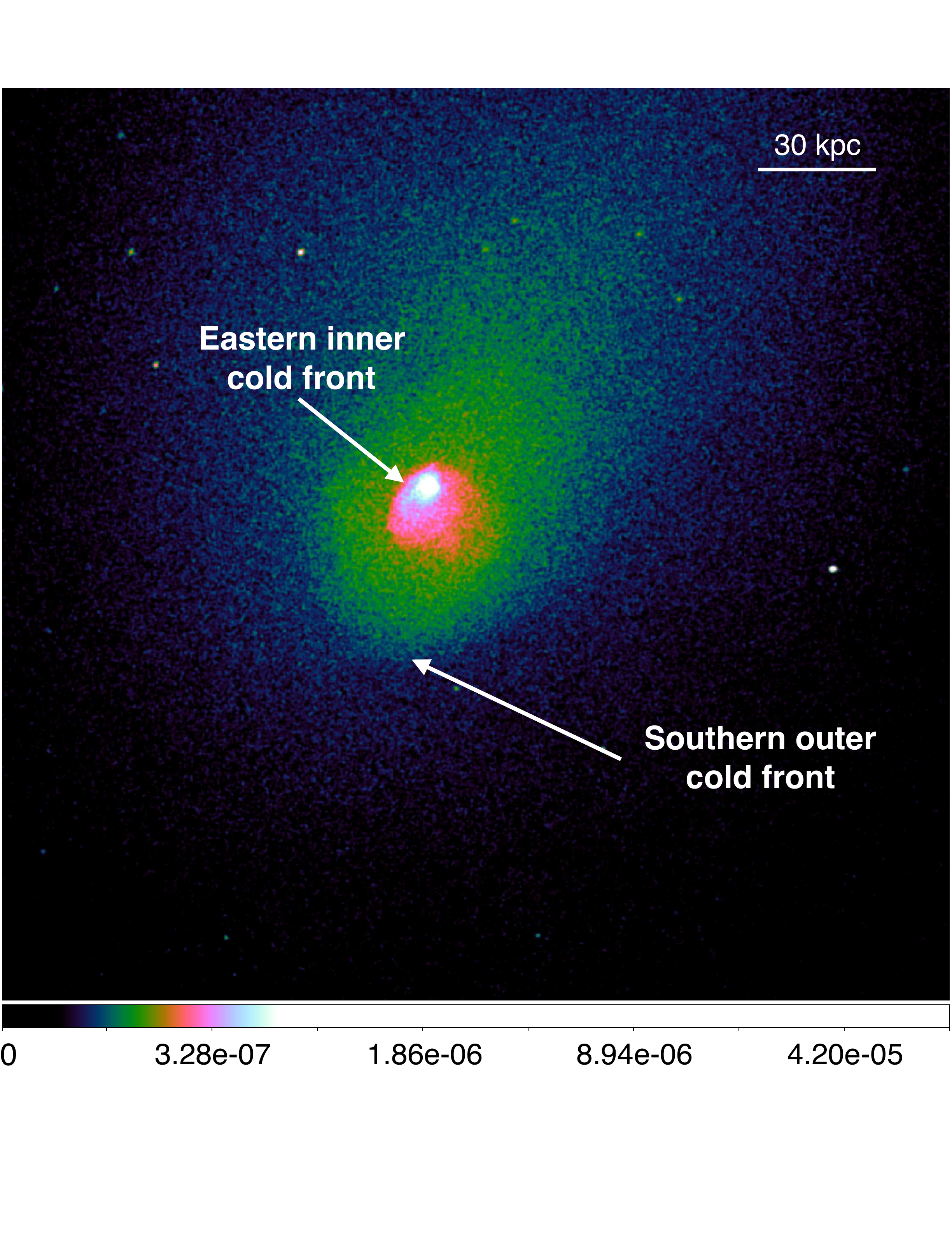}
\end{minipage}
\begin{minipage}{0.45\textwidth}
\vspace{-1.85cm}
\hspace{-0.5cm}
\includegraphics[width=1.25\textwidth,clip=t,angle=0.]{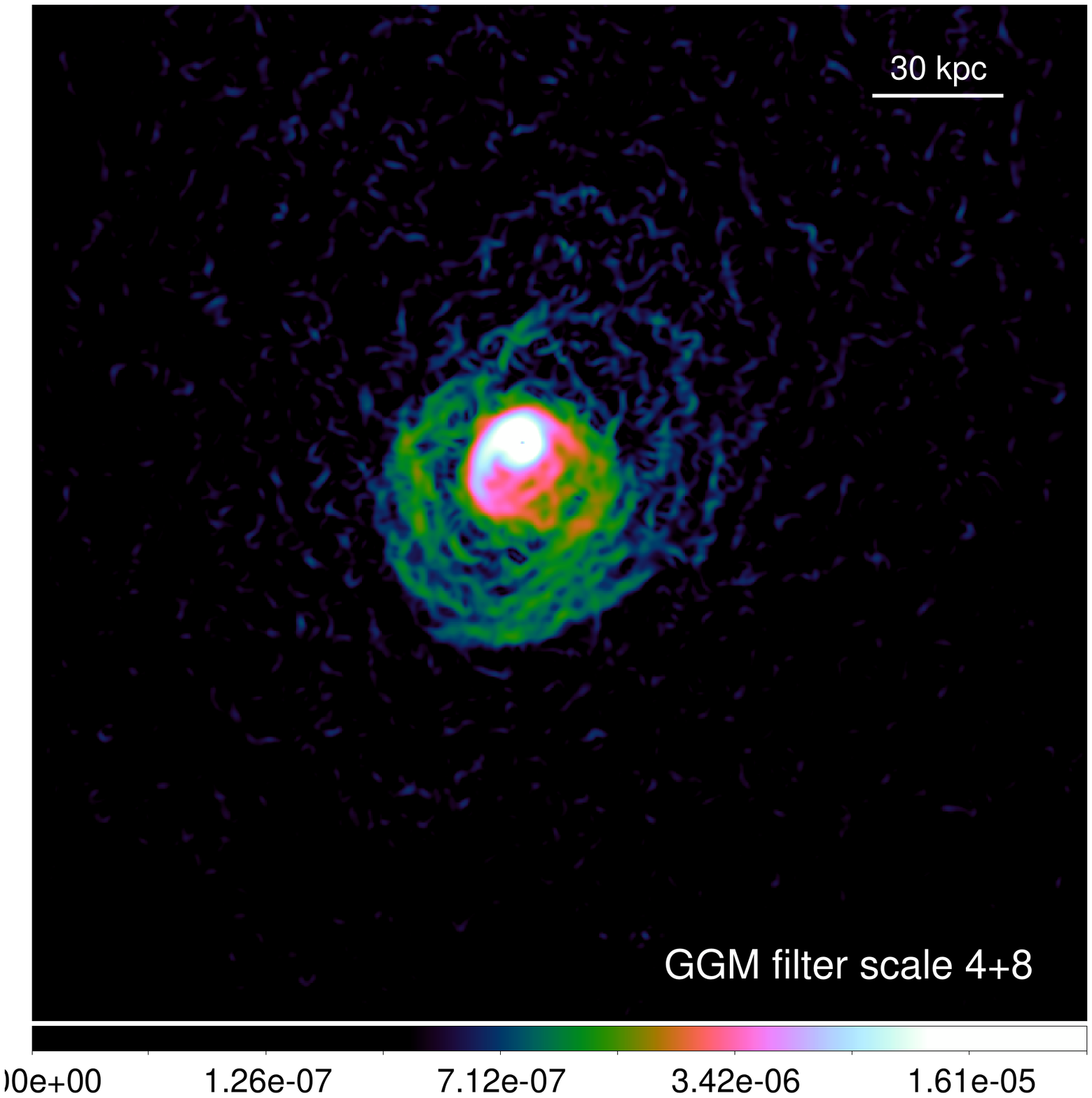}
\end{minipage}
\vspace{-1.5cm}
\caption{The same \chandra\ image as Fig.~\ref{large_IM}, zoomed in on the central part of the cluster (left panel). The right panel shows the GGM filtered image, using $\sigma=4$ and 8 pixels (a pixel corresponds to 0.984 arcsec), which highlights multiple edges and the disturbed, broken-up cold front to the southeast of the cluster core. } 
\label{m_IM}
\end{figure*}

\label{analysis}
\begin{figure*}
\begin{minipage}{0.45\textwidth}
\vspace{-2.1cm}
\hspace{-0.3cm}
\includegraphics[width=1.07\textwidth,clip=t,angle=0.]{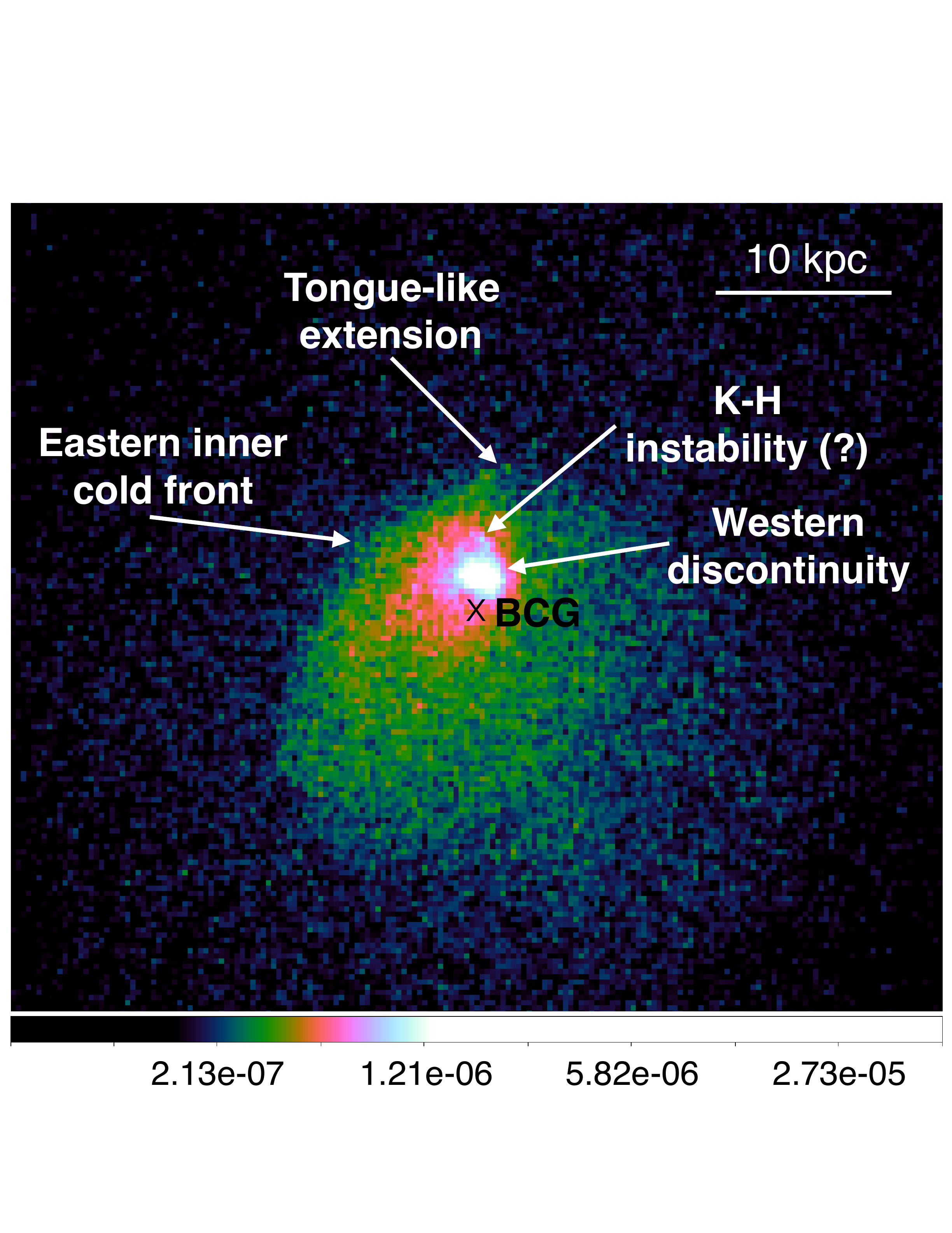}
\end{minipage}
\begin{minipage}{0.45\textwidth}
\vspace{-1.55cm}
\hspace{-0.5cm}
\includegraphics[width=1.25\textwidth,clip=t,angle=0.]{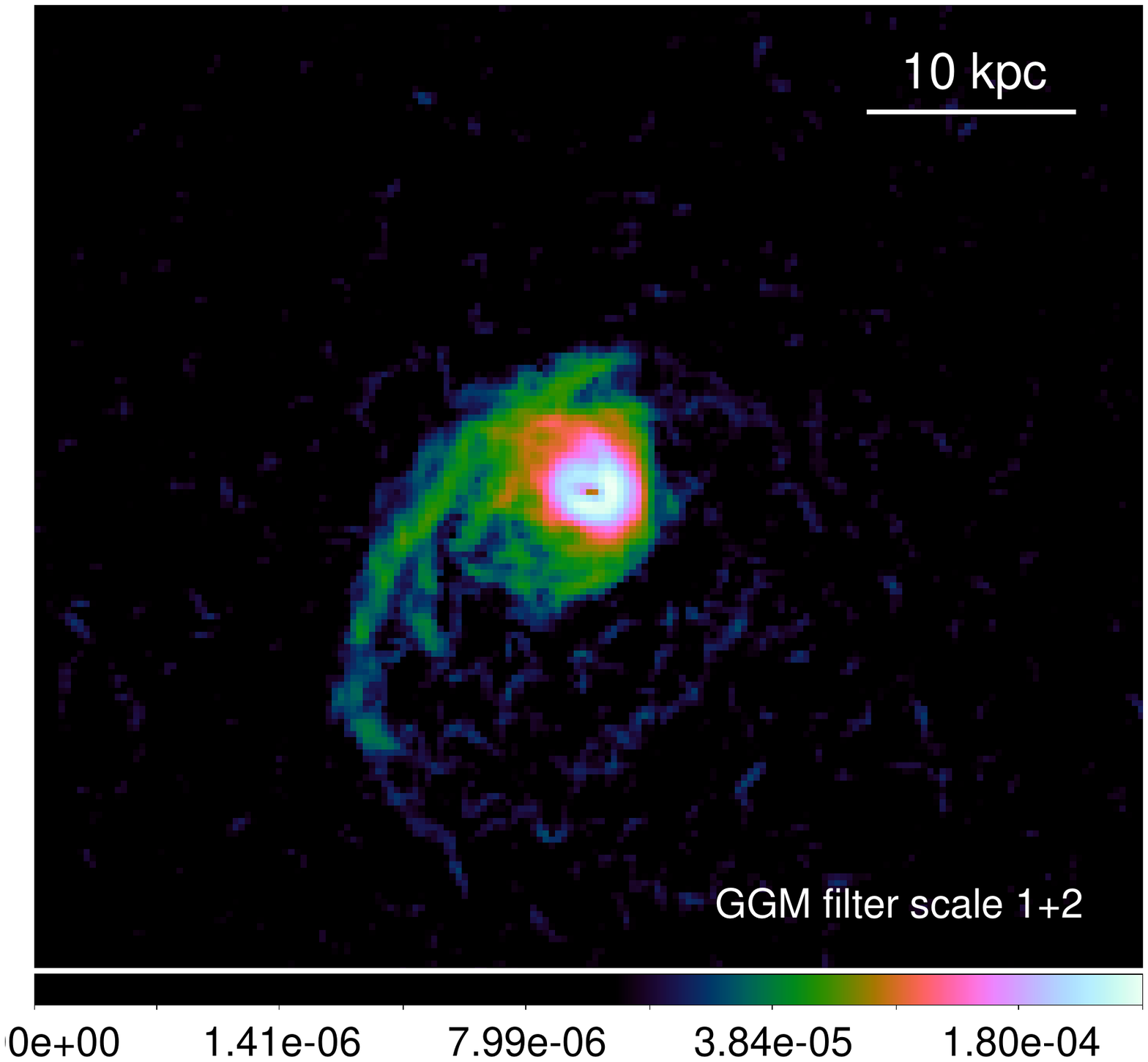}
\end{minipage}
\vspace{-2.5cm}
\caption{The same \chandra\ image as Fig.~\ref{large_IM}, zoomed in on the very central part of the cluster core (left panel). The right panel shows the GGM filtered image, using $\sigma=1$ and 2 pixels, which highlights the remarkably sharp eastern inner cold front with a `tongue'-like extension to the north, and the western discontinuity with a possible Kelvin-Helmholtz roll. The X-ray surface brightness peak is offset from the center of the BCG by about 2.2~kpc. } 
\label{s_IM}
\end{figure*}

\subsubsection{Image analysis}

Background-subtracted images, using both the native resolution of the {\it Chandra} CCDs and using $2\times2$ pixel binning, were created in 13 narrow energy bands (0.6--0.8~keV, 0.8--1.0~keV, 1.0--1.2~keV, 1.2--1.5~keV, 1.5--2.0~keV, 2.0--2.5~keV, 2.5--2.75~keV, 2.75--3.0~keV, 3.0--3.5~keV, 3.5--4.0~keV, 4.0--6.0~keV, 6.0--7.0~keV, 7.0--7.5~keV), spanning 0.6--7.5 keV. The narrow-band images were flat fielded with respect to the median energy for each image and then co-added to create the X-ray images shown in Fig.~\ref{large_IM}. Identification of point sources was performed using the {\texttt{CIAO}} task {\texttt{WAVDETECT}}. The point sources were excluded from the subsequent analysis. 

To look for sub-structure in the cluster, we removed the underlying large-scale surface brightness gradient by fitting an elliptical double beta-model to the image using the {\texttt{Sherpa}} package, and subsequently divided the image by the best-fit model (see the right panel of Fig.~\ref{large_IM}).  

To highlight the sharp features in the image, we also used a Gaussian Gradient Magnitude (GGM) filter \citep[see][]{sanders2016}. The GGM filter uses Gaussian derivatives to determine the magnitude of surface brightness gradients in the image. We applied the GGM filter to the exposure-corrected, background subtracted image, from which point sources have been removed, for various Gaussian widths, using $\sigma=1$, 2, 4, and 8 pixels (a pixel corresponds to 0.984 arcsec).
Regions with large gradients, such as discontinuities, are bright, while flatter surface brightness areas are darker (see the right panel of Fig.~\ref{m_IM}).

To study the detailed properties of the surface brightness discontinuities in our X-ray images, we extracted surface brightness profiles at the native resolution of the {\it Chandra} CCD detectors, in 0.492 arcsec bins. We optimized the sharpness of the surface brightness edges in the profiles by aligning our circular annuli with the discontinuities. We fit the surface brightness profiles by projecting a spherically symmetric discontinuous, broken power-law density distribution \citep[see e.g.][]{owers2009}. The fitting was performed using the {\texttt{PROFFIT}} package \citep{eckert2011} modified by \citet{ogrean2015}. In order to place upper limits on the widths of the edges, we included in our models Gaussian smoothing as an additional free parameter.

\subsubsection{Spectral analysis}

Individual regions for the 2D spectral mapping are determined using the contour binning algorithm \citep{sanders2006b}, which groups neighboring pixels of similar surface brightness until a desired signal-to-noise ratio threshold is met. The Ophiuchus cluster is a very hot system, with $kT\sim10$~keV \citep{matsuzawa1996,fujita2008b,million2010}, which makes its temperature and metallicity measurement relatively difficult (at such high temperatures the cut-off of the thermal bremsstrahlung spectrum is outside of the energy window of the telescope), requiring a large number of counts. Therefore, we adopt a relatively high signal-to-noise ratio of 200 (approximately 40,000 counts per region), which provides approximately 2 per cent precision for density, 4 per~cent precision for temperature, 12 per cent precision for metallicity, and 3 per cent precision for the line-of-sight absorption column density measurements, respectively. To obtain a high resolution 2D spectral map of the cooling core, we also produced a map of the innermost $r=10$~kpc region with a signal-to-noise of 50 (approximately 2500 counts per region), which provides 3--9 per~cent precision for temperature, and 18--25 per cent per cent for metallicity. The fractional errors are the smallest for the lowest temperature regions and grow bigger with increasing temperature. We model the spectra extracted from each spatial region with the {\texttt{SPEX}} package \citep{kaastra1996} in the 0.6--7.5 keV band. The spectrum for each region is fitted with a model consisting of an absorbed single-phase plasma in collisional ionization equilibrium. The Ophiuchus cluster lies close to the plane of our Galaxy, and therefore its line-of-sight absorption column density is high, with $N_{\rm H}$, according to the Leiden/Argentine/Bonn radio survey of \ion{H}{i}, of $N_{\rm H}=2\times10^{21}$~cm$^{-2}$ \citep{kalberla2005}. Radio surveys of \ion{H}{i} usually underestimate the total absorption for systems with such high $N_{\rm H}$, because they are not sensitive to molecular and dust absorption. Therefore, the $N_{\rm H}$ has to be a free parameter in our model. Other free parameters include the spectral normalization (emission measure), temperature, and metallicity. 

The deprojected spectra are obtained using the {\texttt{PROJCT}} model in {\texttt{XSPEC}} and the {\texttt{APEC}} plasma model \citep{smith2001,foster2012}. The line-of-sight absorption column density is a free parameter in the fit and for a given azimuth it is assumed to have the same value in all annuli, except the innermost region where it is allowed to vary independently. Metallicities are given with respect to the proto-Solar abundances by \citet{grevesse1998}.

\subsection{Radio JVLA data}

We observed the Ophiuchus cluster with the Jansky Very Large Array (JVLA) in the L band for four hours on 2015 July 30. The observation was taken in the A
configuration, resulting in a restoring FWHM beam size of $1.17\arcsec\times2.48\arcsec$. We spent 16 minutes on 3C286 to calibrate the flux, 22 minutes on J1734+0926 to calibrate the polarization leakage, 18 minutes on J1714-2514 to calibrate the phase, and a total of 150 minutes on source. The total bandpass was 1 GHz.

The data analysis was performed using the {\texttt{CASA}} package \citep{mcmullin2007}. After the flux and phase calibrations were transferred to the
source and RFI excluded, we used the {\texttt{CLEAN} routine to image the cluster. We used a Briggs weighting of ${\rm Robust}=0.7$ to increase the image sensitivity, while still keeping the high resolution to detect any possible extended structure near the BCG. We created the images fitting for both spectral index and correcting for the primary beam attenuation. The final image has an rms of 10 $\mu$Jy.

\begin{figure*}
\vspace{-1.5cm}
\begin{minipage}{0.33\textwidth}
\hspace{-1cm}
\includegraphics[width=1.18\textwidth,clip=t,angle=0.]{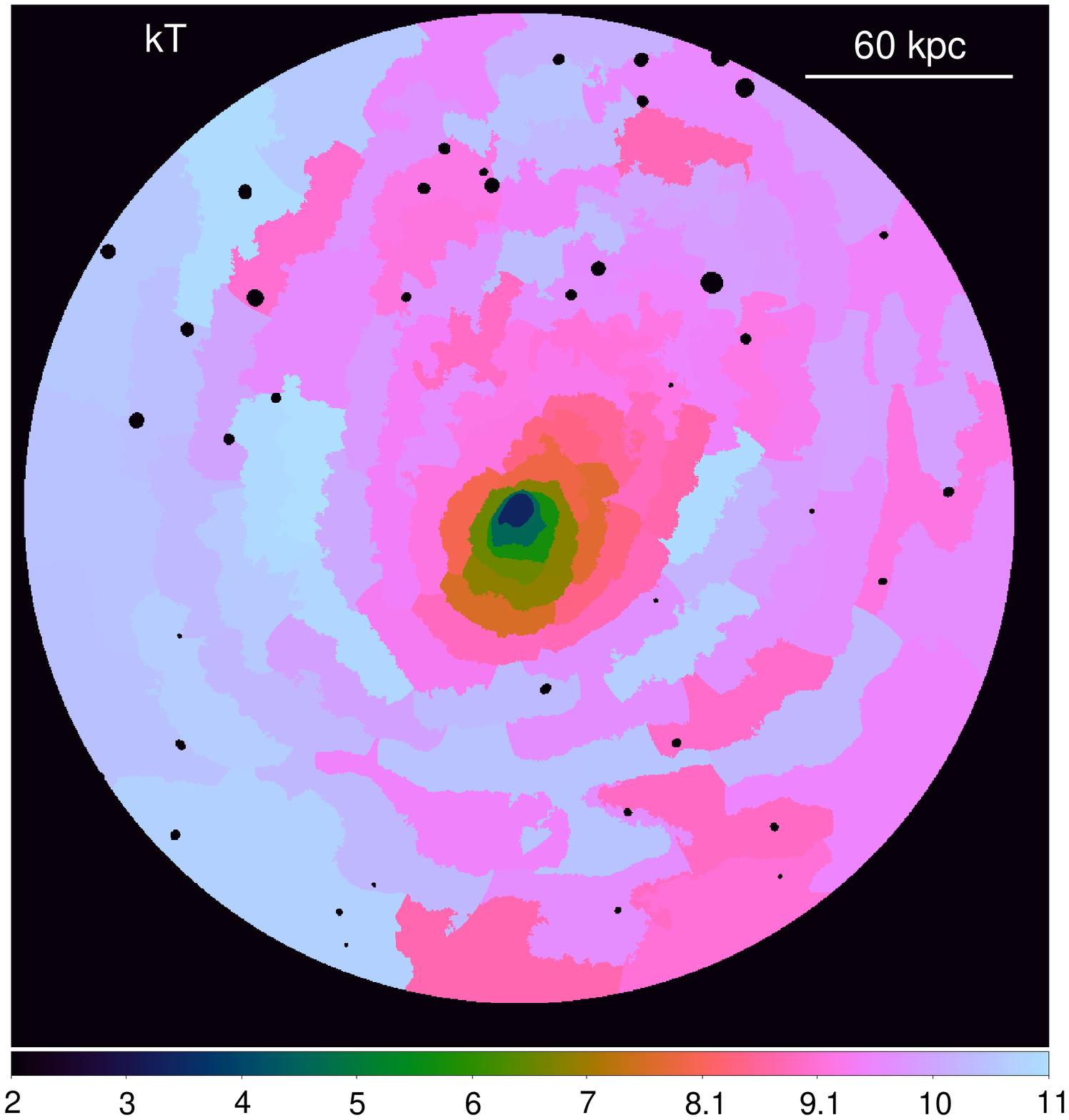}
\end{minipage}
\begin{minipage}{0.33\textwidth}
\hspace{-0.5cm}\includegraphics[width=1.18\textwidth,clip=t,angle=0.]{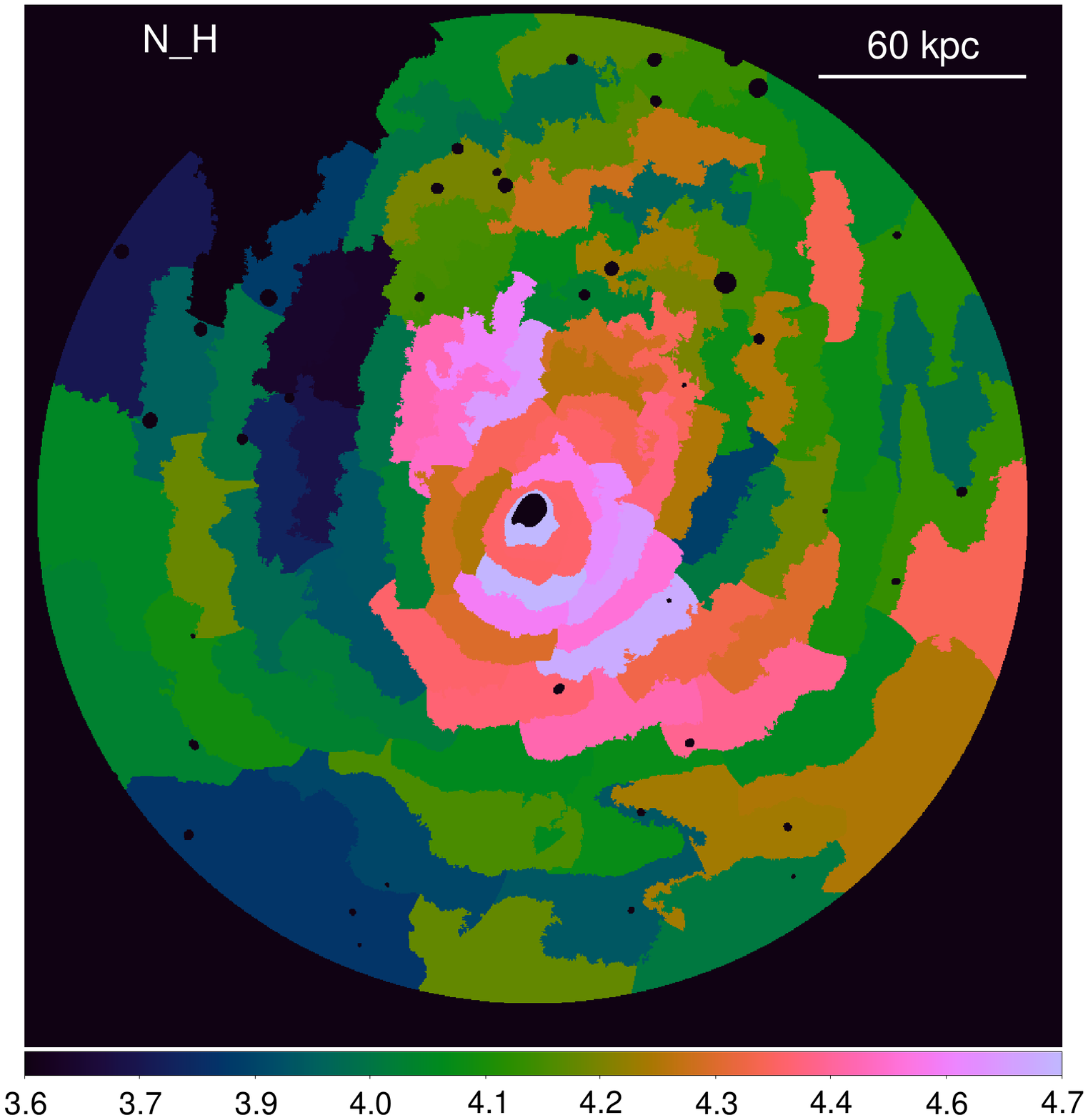}
\end{minipage}
\begin{minipage}{0.33\textwidth}
\includegraphics[width=1.18\textwidth,clip=t,angle=0.]{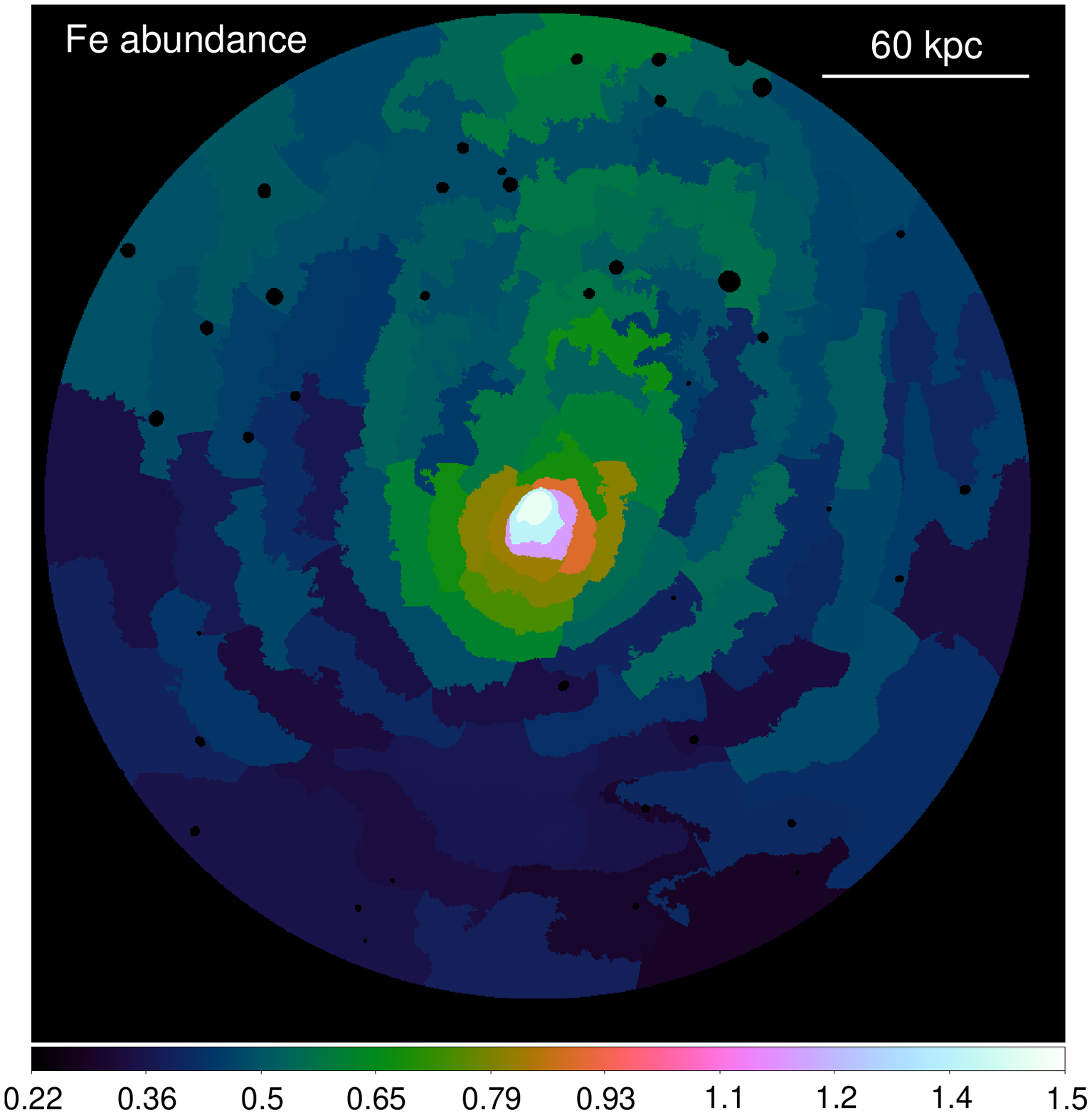}
\end{minipage}
\vspace{-1.5cm}
\caption{2D maps of the best fit projected temperature, line of sight absorbing hydrogen column density, and iron abundance. Each spatial region contains 40,000 net counts (S/N of 200). The units of temperature, absorbing column density, and Fe abundance are keV, $10^{21}$~cm$^{-2}$, and Solar \citep{grevesse1998}. The fractional $1\sigma$ statistical errors are $\sim 4$ per~cent for temperature, and $\sim12$ per cent for metallicity. Point sources were excluded from the regions shown as black circles. } 
\label{maps}
\end{figure*}

\begin{figure*}
\vspace{-1.5cm}
\begin{minipage}{0.33\textwidth}
\hspace{-1cm}
\includegraphics[width=1.18\textwidth,clip=t,angle=0.]{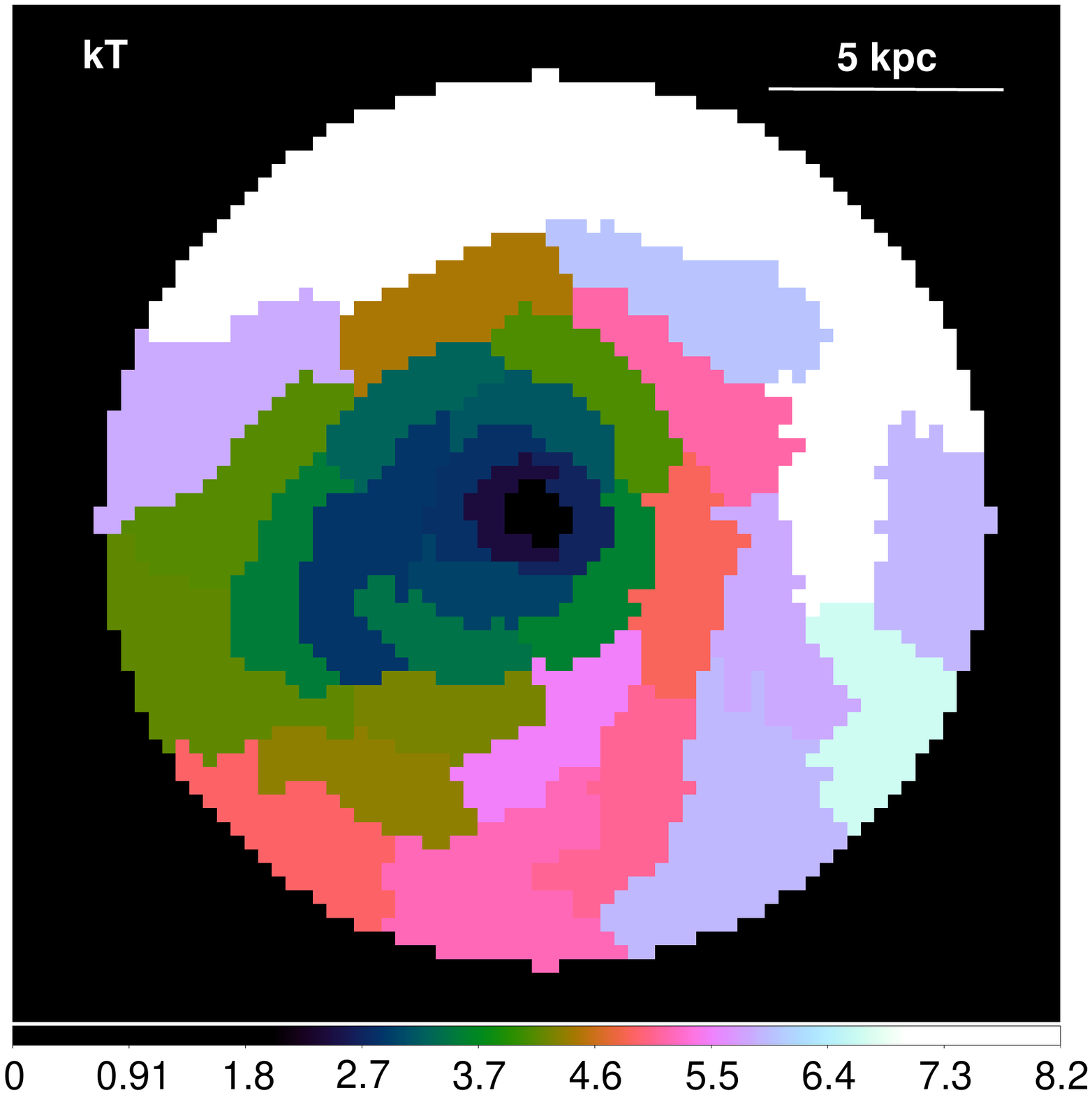}
\end{minipage}
\begin{minipage}{0.33\textwidth}
\hspace{-0.5cm}\includegraphics[width=1.18\textwidth,clip=t,angle=0.]{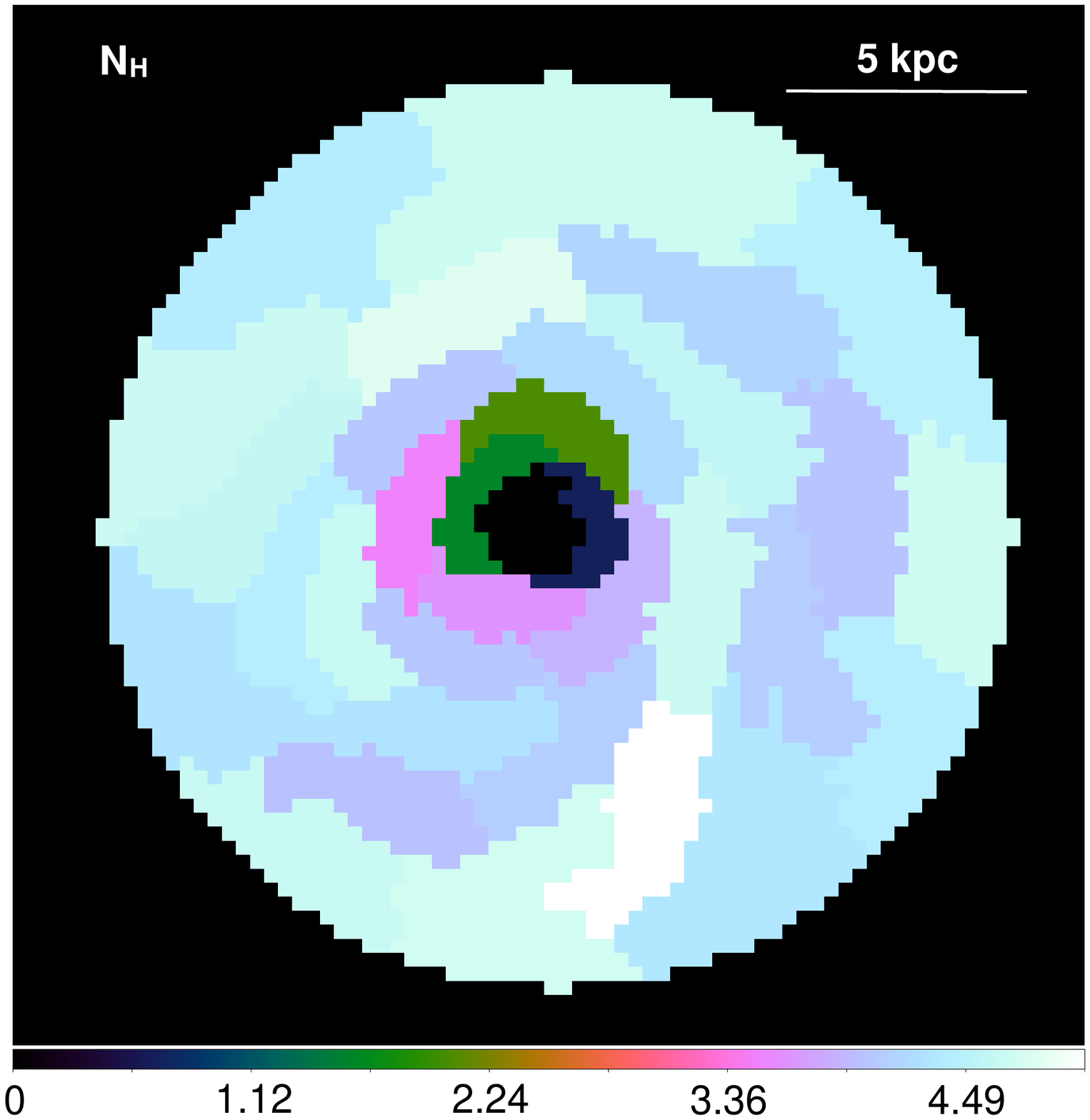}
\end{minipage}
\begin{minipage}{0.33\textwidth}
\includegraphics[width=1.18\textwidth,clip=t,angle=0.]{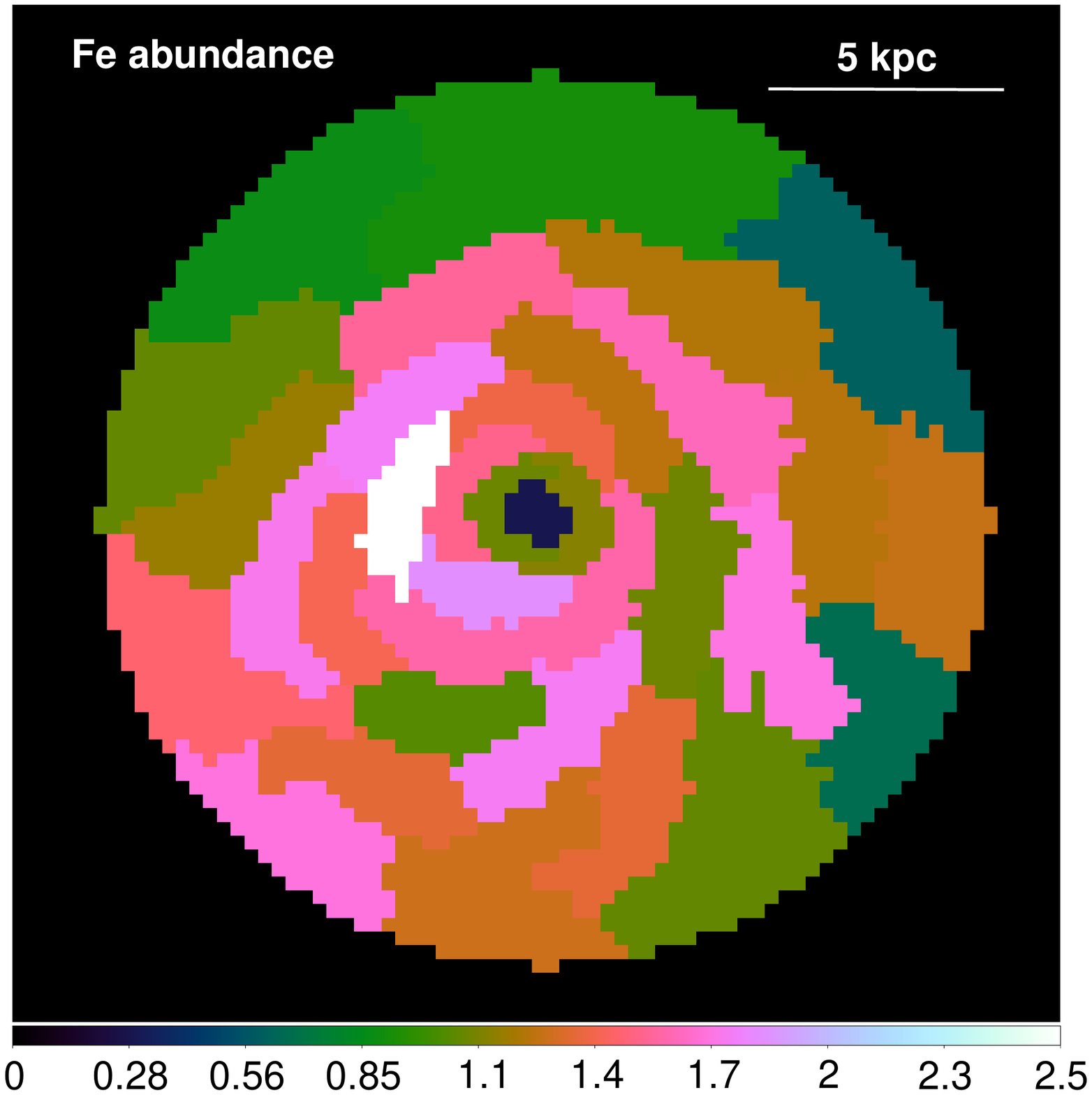}
\end{minipage}
\vspace{-1.5cm}
\caption{The 2D maps zoomed-in on the innermost $r=10$~kpc core of the Ophiuchus cluster show the best fit projected temperature (left panel), line of sight absorbing hydrogen column density (central panel), and iron abundance (right panel) for spatial regions with 2500~counts (S/N of 50).  The units of temperature, absorbing column density, and Fe abundance are keV, $10^{21}$~cm$^{-2}$, and Solar \citep{grevesse1998}. The fractional $1\sigma$ statistical errors are 3--9 per~cent for temperature, and 18--25 per cent per cent for metallicity. The fractional errors are the smallest for the lowest temperature regions and grow larger with increasing temperature. } 
\label{maps_HR}
\end{figure*}

\label{analysis}
\begin{figure*}
\vspace{-1.5cm}
\begin{minipage}{0.45\textwidth}
\hspace{-1cm}
\includegraphics[width=1.25\textwidth,clip=t,angle=0.]{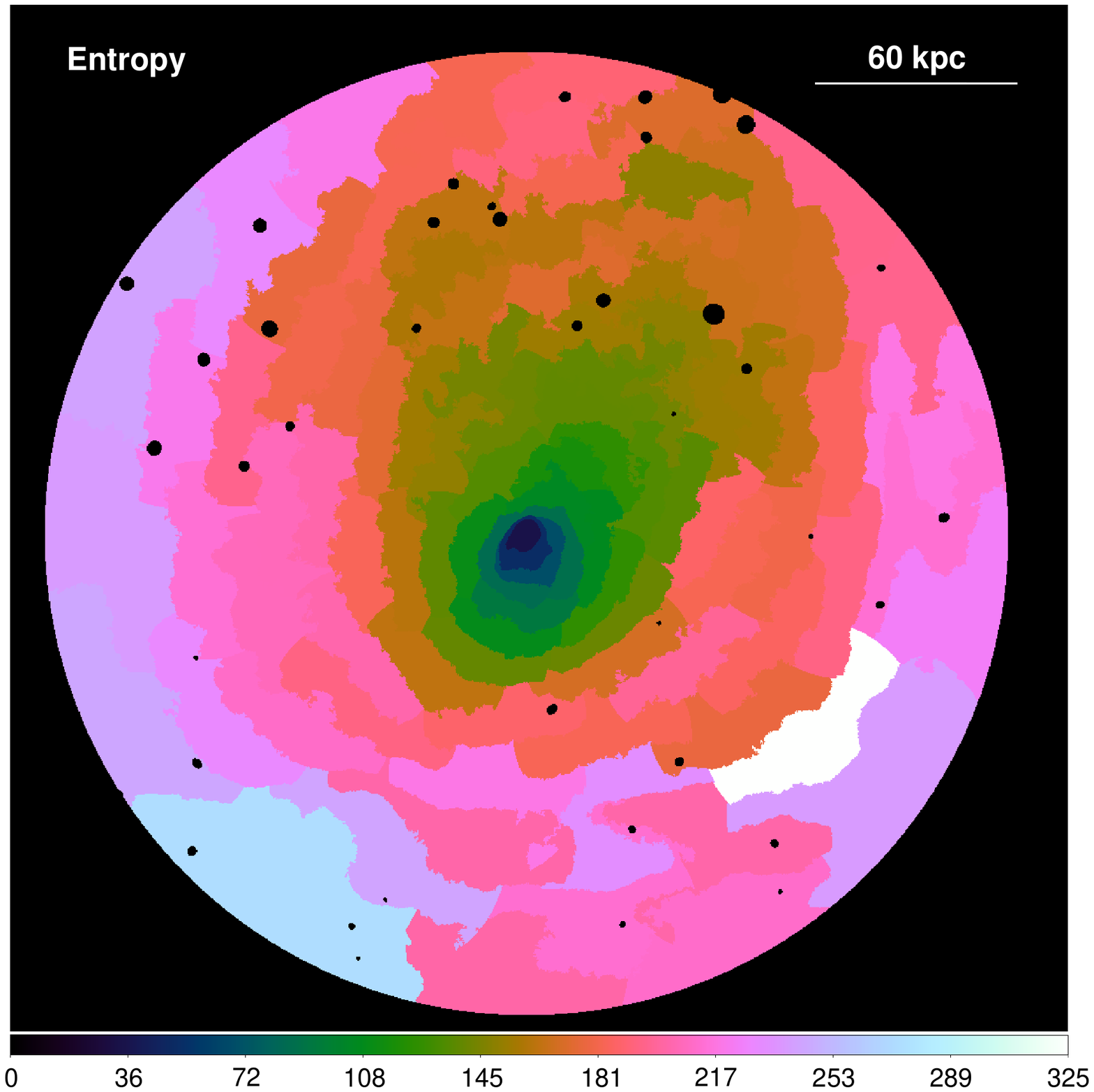}
\end{minipage}
\begin{minipage}{0.45\textwidth}
\includegraphics[width=1.25\textwidth,clip=t,angle=0.]{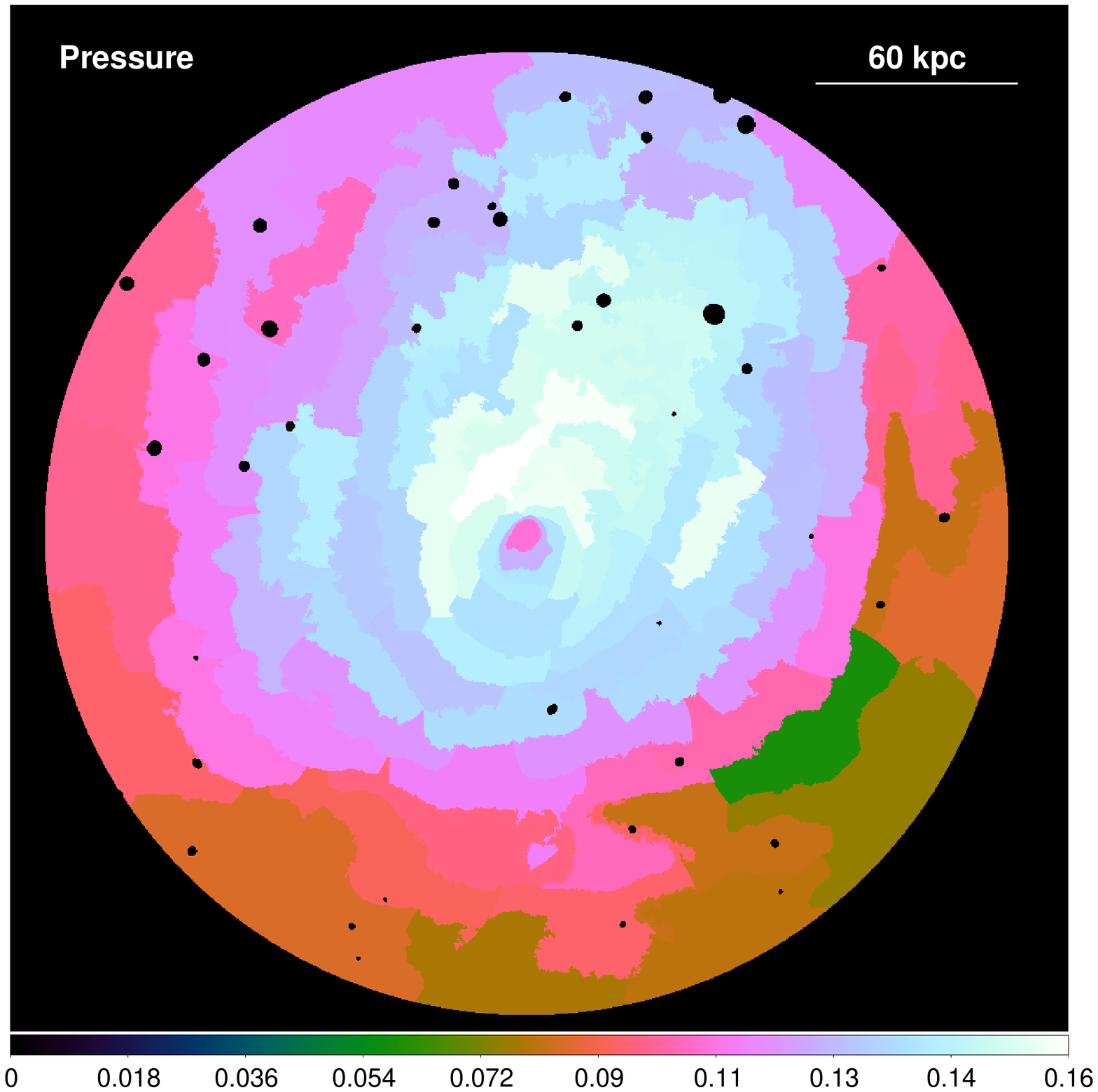}
\end{minipage}
\vspace{-1.5cm}
\caption{2D maps of the projected entropy and pressure. To study their azimuthal distribution, we assume the gas uniformly distributed along the line-of-sight depth of $l = 100$ kpc over the entire field of view. The units of the entropy and pressure are keV~cm$^2(l/\rm{100~kpc})^{1/3}$ and keV~cm$^{-3}(l/\rm{100~kpc})^{1/2}$, respectively. The fractional $1\sigma$ statistical errors are about 5 per cent.} 
\label{thermo_maps}
\end{figure*}

\begin{figure*}
	\begin{minipage}{0.32\textwidth}
		\includegraphics[width=1.2\textwidth,clip=t,angle=0.]{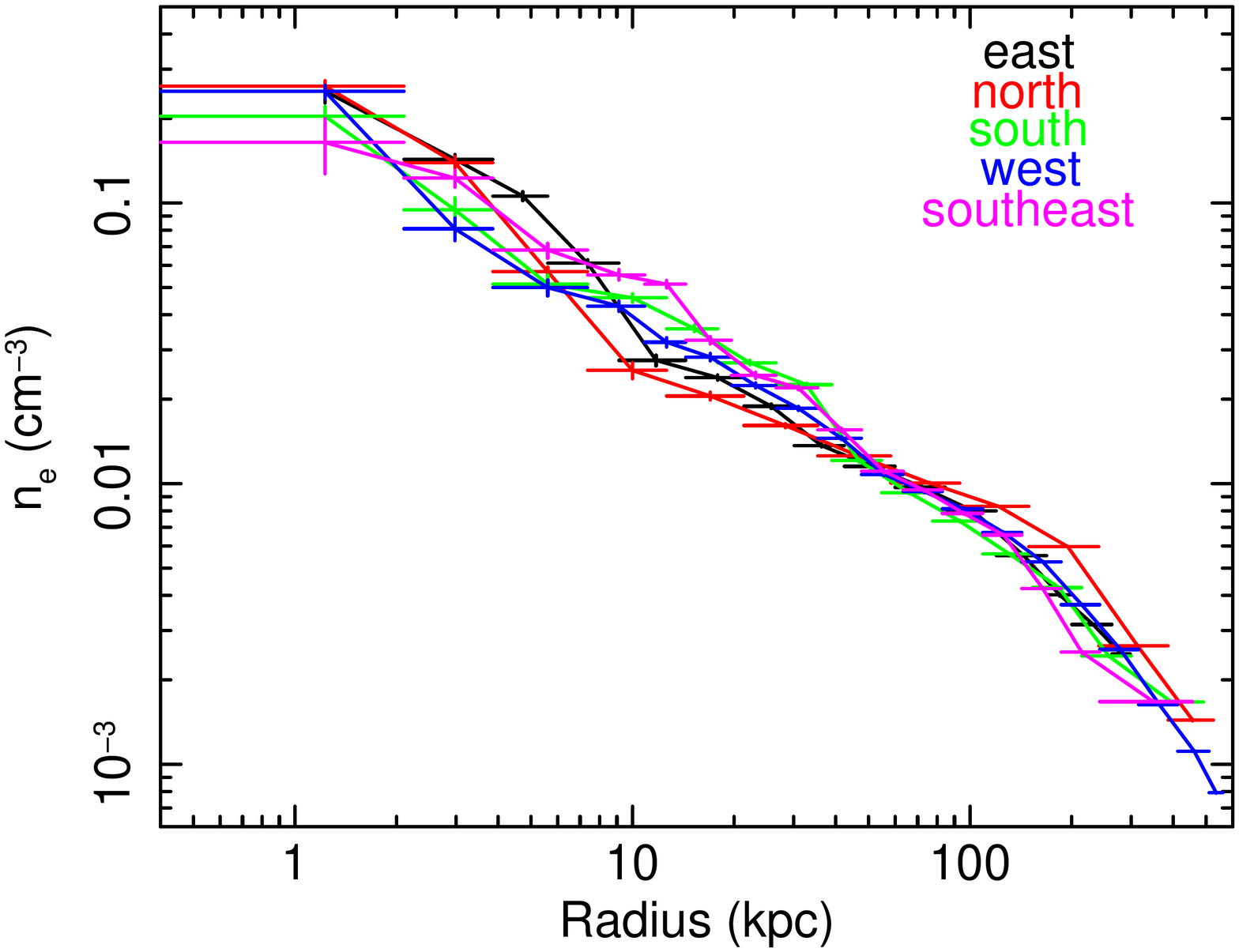}
	\end{minipage}
	\begin{minipage}{0.32\textwidth}
		\includegraphics[width=1.2\textwidth,clip=t,angle=0.]{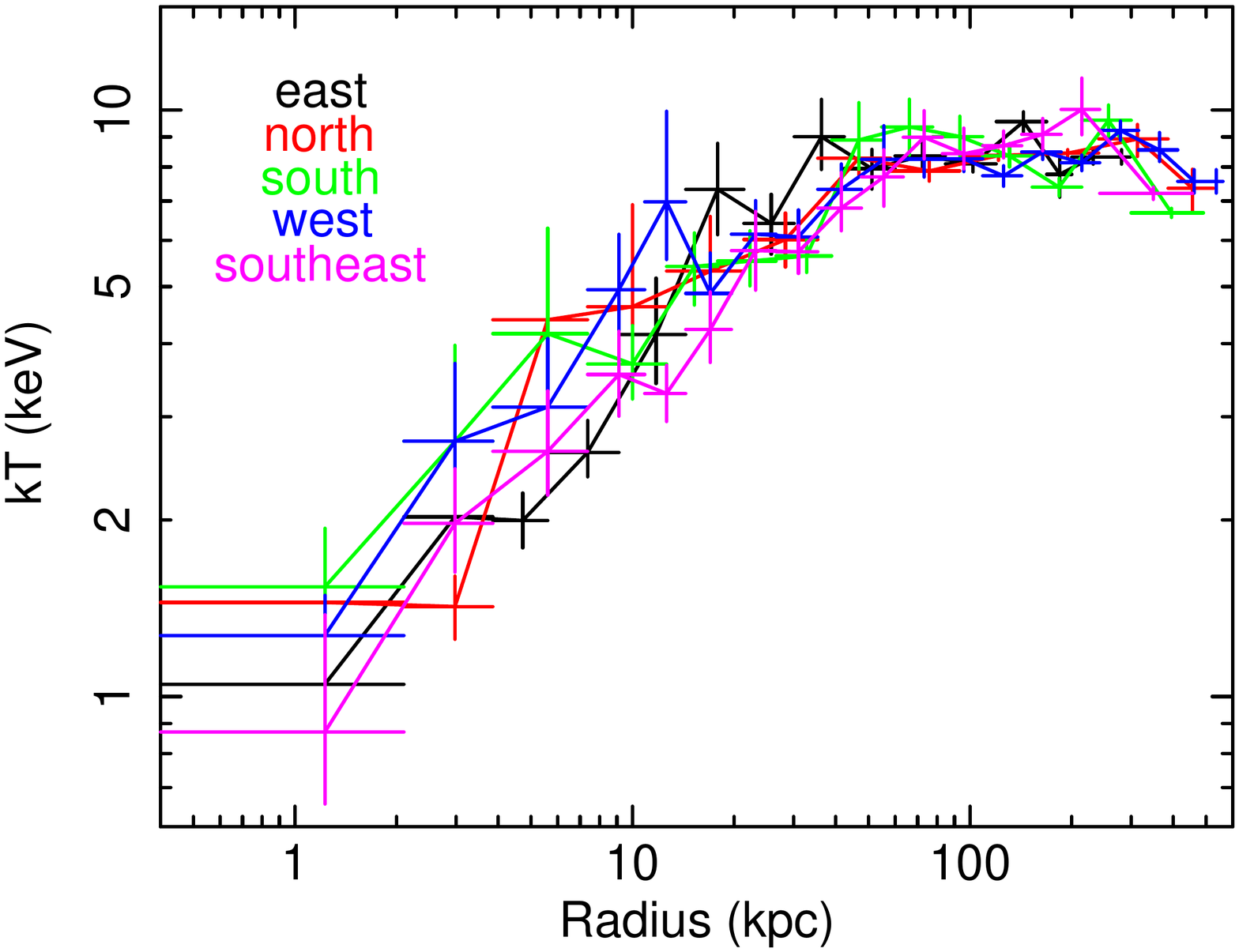}
	\end{minipage}
	\begin{minipage}{0.32\textwidth}
		\includegraphics[width=1.2\textwidth,clip=t,angle=0.]{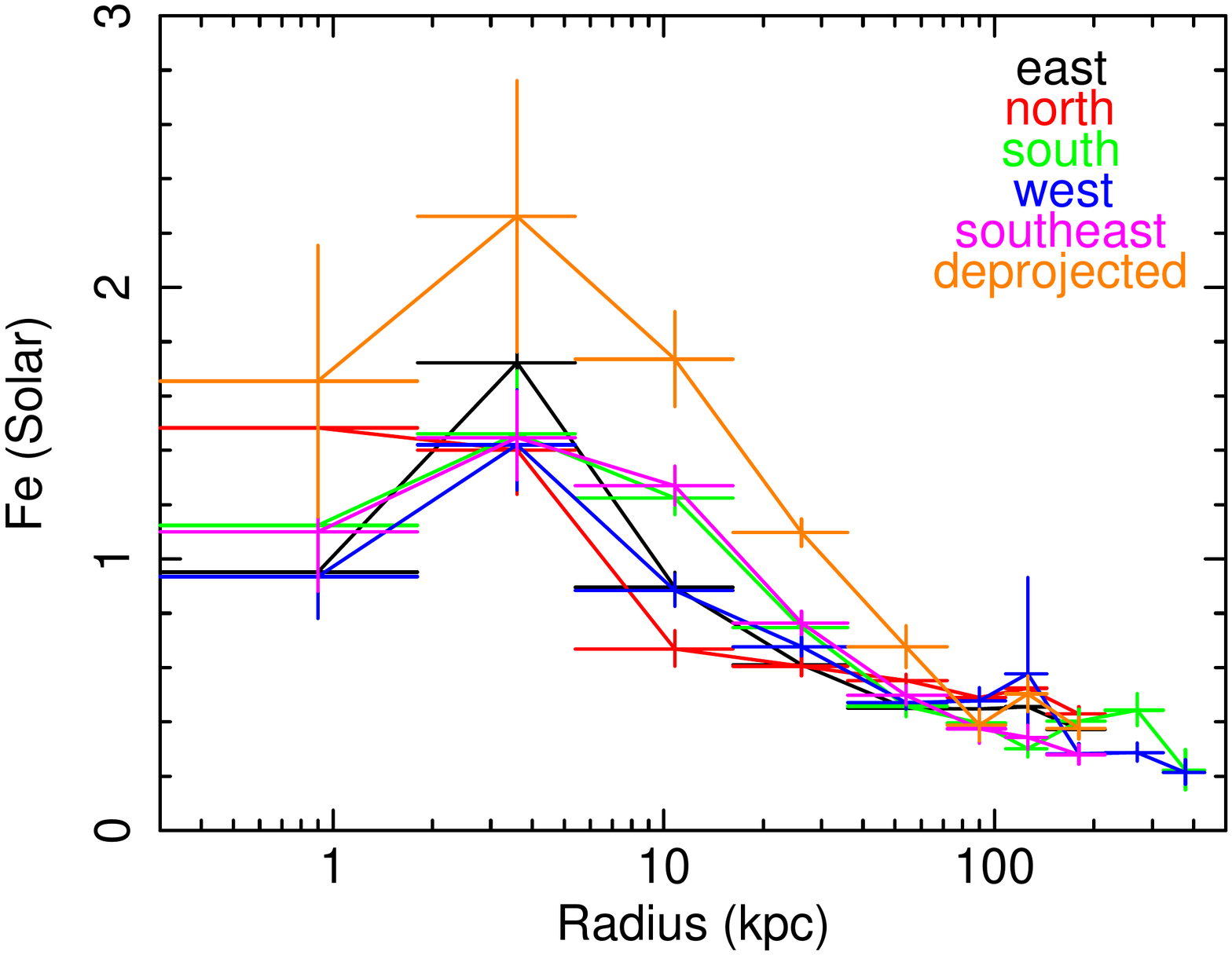}
	\end{minipage}
	\caption{Radial profiles of the deprojected electron density and temperature distribution determined along five different azimuths (the azimuth angles are measured counterclockwise from the west): west (305--30 degrees), north (30--110), east (110--205), southeast (205--260), and south (260--305). For the iron abundance distribution, in the right panel, we show the projected measurements and the azimuthally averaged deprojected profile. } 
	\label{profiles1}
\end{figure*}

\begin{figure*}
	\begin{minipage}{0.45\textwidth}
		\includegraphics[width=1.2\textwidth,clip=t,angle=0.]{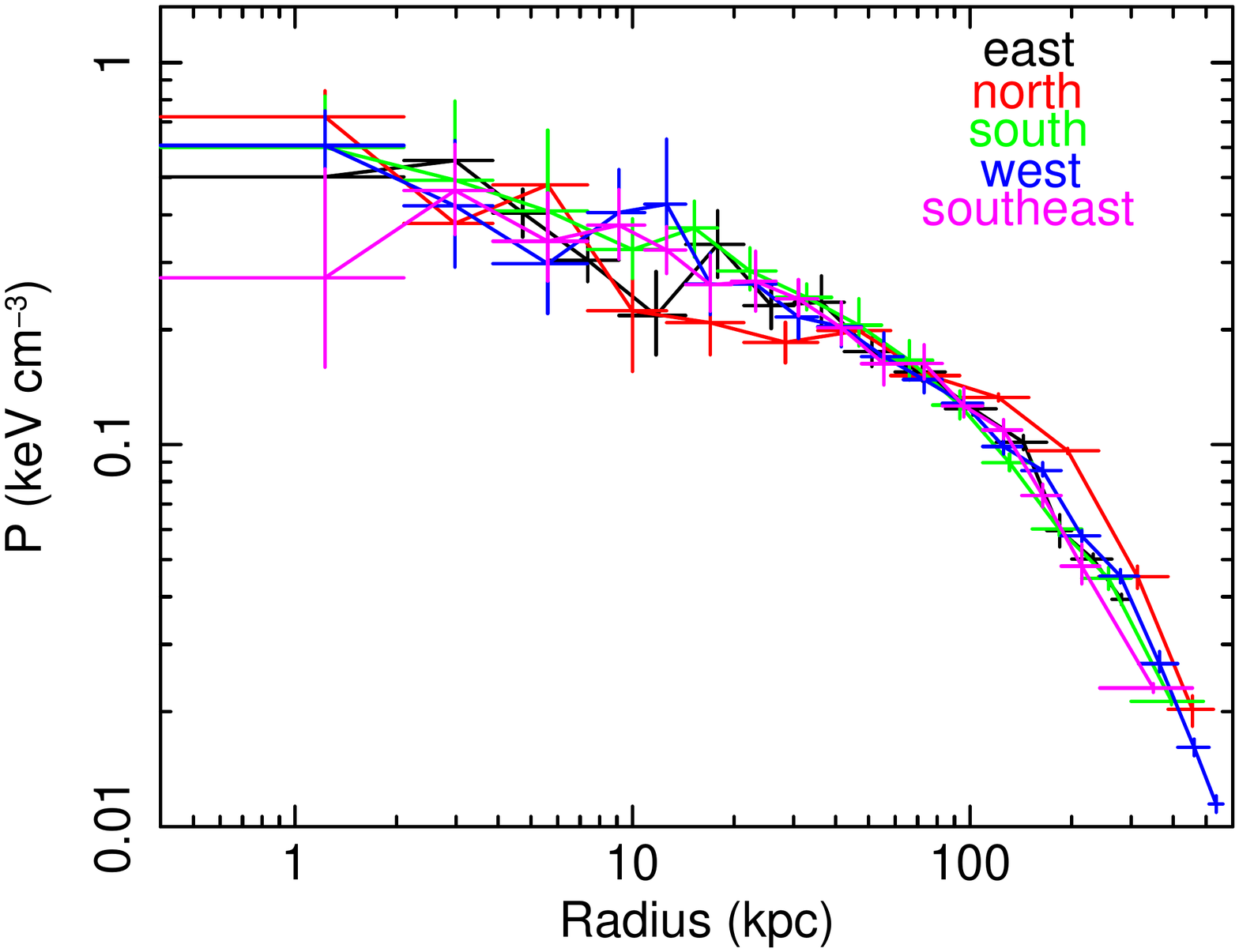}
	\end{minipage}
	\begin{minipage}{0.45\textwidth}
		\includegraphics[width=1.2\textwidth,clip=t,angle=0.]{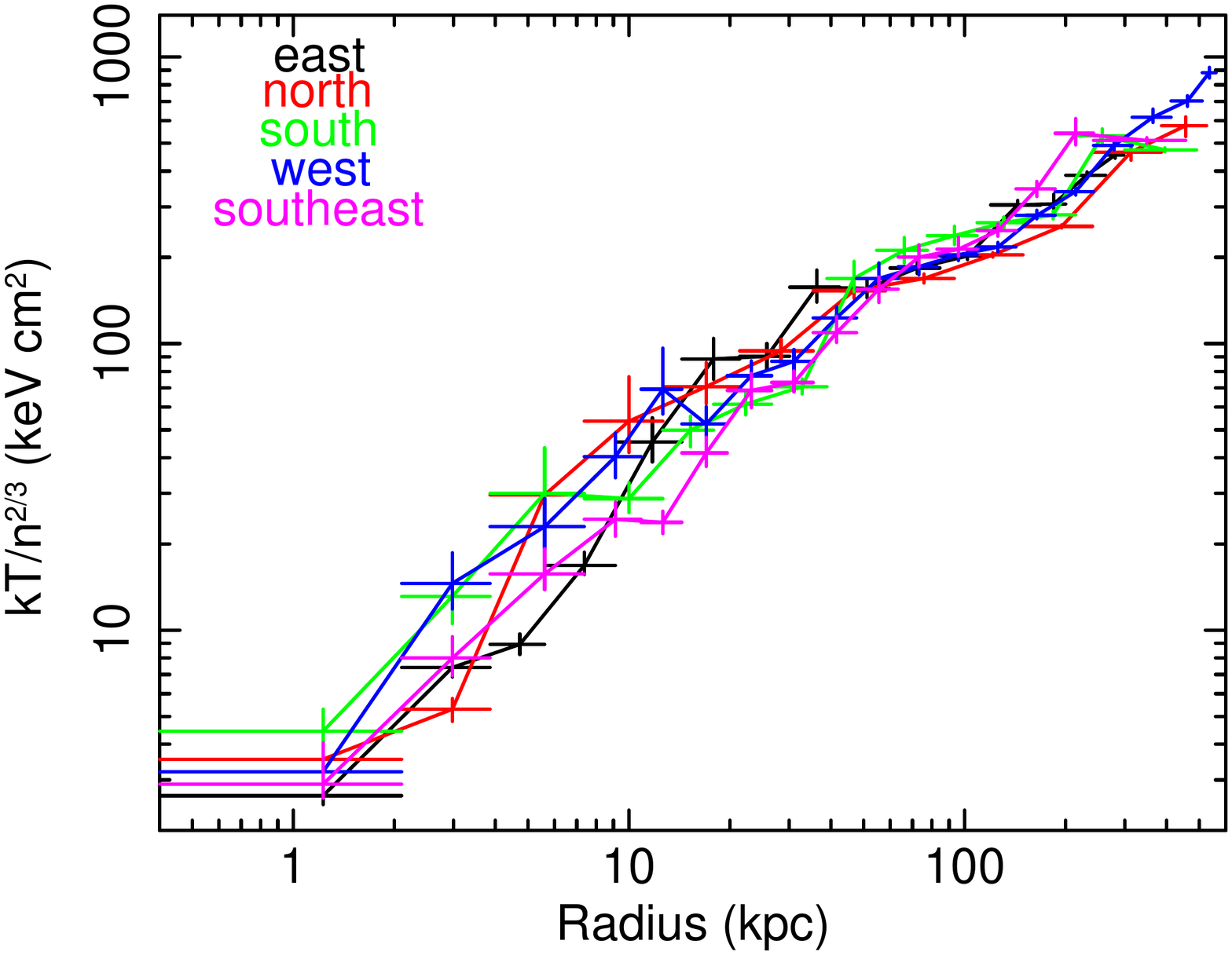}
	\end{minipage}
	\caption{Using the electron density $n_{\rm e}$ and temperature $kT_{\rm e}$ profiles shown in Fig.~\ref{profiles1}, we determine the radial profiles of the total pressure ($P=nkT$; left panel) and entropy ($K=kT_{\rm e}/n^{2/3}_{\rm e}$; right panel).} 
	\label{profiles2}
\end{figure*}

\section{Results}
\label{results}
\subsection{X-ray imaging}
\label{imaging}

The Ophiuchus cluster shows a strongly centrally peaked X-ray surface brightness distribution, elongated north-northwestward (see the left panel of Fig.~\ref{large_IM}). By fitting an elliptical double beta-model to the image, with the cluster center as a free parameter, we find that the center of the large scale emission distribution is offset by 38~kpc to the north-northwest of the central emission peak and the brightest cluster galaxy (BCG). In the residual image, obtained by dividing our {\it Chandra} image by the best fit elliptical double beta-model (see the right panel of Fig.~\ref{large_IM}), the tail-like excess emission extending north-northwestward practically disappears, indicating that this apparent excess emission is the result of the bright cluster core being displaced southward of the center of the global emission distribution. The residual image reveals a surface brightness excess at the projected distance of $r\sim280$~kpc to the south of the BCG, which coincides with some bright galaxies identified in 2MASS images. Its southern edge has a hint of a sharp bow-shock shaped morphology. The residual image also reveals a surprising, sharp surface brightness discontinuity that is curved away from the core, at $r\sim120$~kpc (3.55 arcmin) to the southeast of the BCG. 
 
The cluster core harbors a series of nested cold fronts \citep[previously discussed in][]{ascasibar2006,million2010} that surround the cool core (see Fig.~\ref{m_IM}). The outermost cold front wraps around the core from the east to the west and extends out to $r\sim43$~kpc in the south. The best fit density discontinuity at its southern part is $n_{\rm j}=1.7\pm0.1$ and the 99 per cent upper limit on its width is 5.3 kpc. Southeast of the cluster core, the surface brightness discontinuity appears blurred and the GGM filtered image highlights that the cold front is broken up. The GGM filtering also highlights multiple edges to the south of the core seen in the {\it Chandra} image. 

The sharpest, most prominent cold front extends from the northeast to the southwest of the BCG at $r\sim4.5$--$10$~kpc, and has a northward `tongue'-like extension (see Fig.~\ref{s_IM}). Its best fit density jump is $n_{\rm j}=2.0\pm0.3$ and the front appears remarkably sharp. We find that the 99 per cent upper limit on its width is 1.5 kpc. 

The compact, bright innermost cluster core is displaced from the center of the BCG by 2.2~kpc in projection \citep[see also][]{hamer2012}. The compact core has a surface brightness discontinuity at its west side and an extension, a possible Kelvin-Helmholtz roll, to the north. In contrast to the other nearby cooling cores, the core of the Ophiuchus cluster does not show the presence of any {\it obvious} X-ray cavities produced by the active galactic nucleus (AGN). However, the presence of very small cavities, with radius of up to 1.5~kpc, to the southeast of the core cannot be ruled out.

\begin{figure*}
\vspace{-6cm}
\begin{minipage}{0.95\textwidth}
\includegraphics[width=1\textwidth,clip=t,angle=0.]{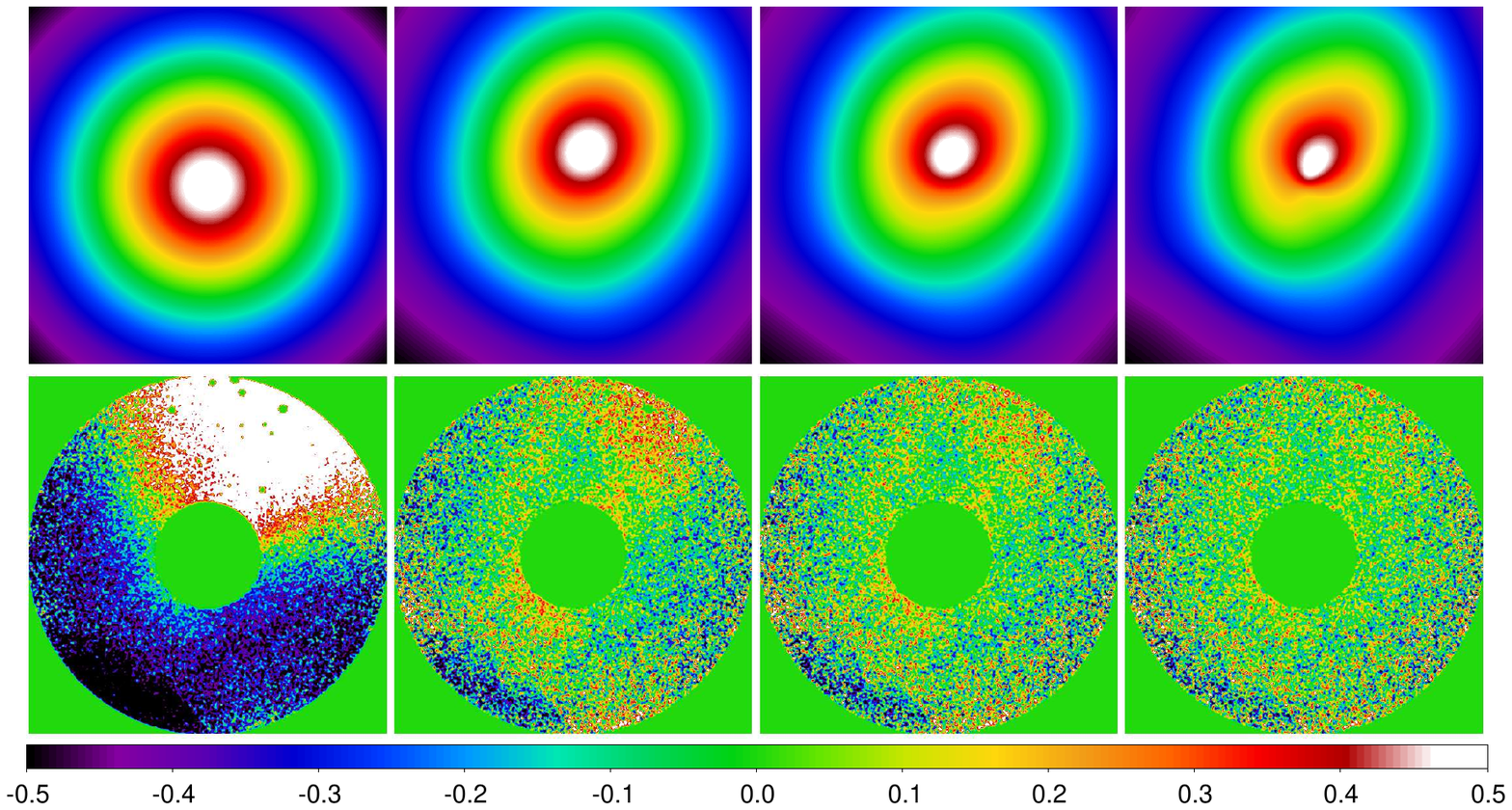}
\end{minipage}
\vspace{-6.7cm}
\caption{{\it Top}: Underlying models of the ``unperturbed'' surface brightness distribution of the Ophiuchus cluster. From left to right: spherically-symmetric $\beta$-model, patched $\beta$-models with $\sigma = 70$~arcsec (removes large-scale asymmetry), $\sigma = 30$~arcsec, and $\sigma = 10$~arcsec \citep[see][for details]{zhuravleva2015}. {\it Bottom:} residual images of the SB fluctuations in the Ophiuchus cluster obtained from the initial image divided by the underlying model on the corresponding upper panel. The smaller the $\sigma$, the smaller the structures included to the model and the less structures remain in the residual image.} 
\label{modelsub}
\end{figure*}

\begin{figure}
\begin{minipage}{0.85\textwidth}
\vspace{-0.75cm}
\hspace{-0.25cm}\includegraphics[width=0.6\columnwidth,clip=t,angle=0.]{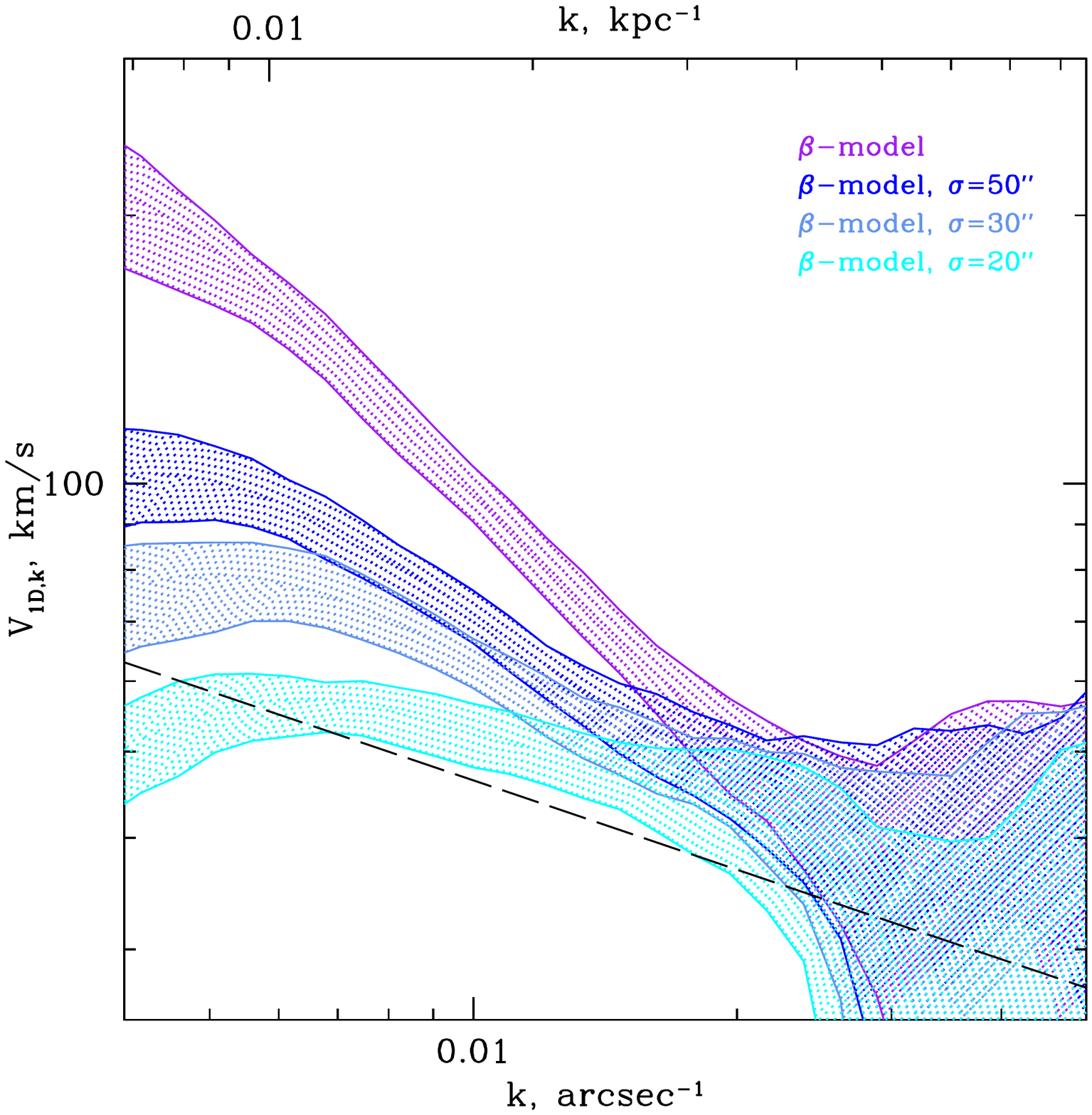}
\end{minipage}
\vspace{-2cm}
\caption{The amplitude of velocity fluctuations in the 1.5--5~arcmin annulus in the Ophiuchus cluster obtained by assuming that the amplitudes of the density and velocity fluctuations are proportional to each other on each scale \citep[see the text and][for detail]{zhuravleva2014,gaspari2014}. The width of each hatched region is the $1\sigma$ statistical uncertainty associated with Poisson noise.   In order to determine the amplitude of density fluctuations, the large scale surface brightness gradient has been modeled with a spherically-symmetric $\beta$-model (top hatched region) and with more flexible models (models patched on scales of $\sigma=70, 30, 10$~arcsec shown in Fig.~\ref{modelsub}). While the top hatched region is clearly affected by the large-scale asymmetry in the cluster, all of the models that account for the asymmetry indicate velocities $V_{\rm{1D,k}}\lesssim100$~km~s$^{-1}$ on scales $\lesssim100$~kpc. The dashed line indicates the expected shape of the  spectrum of Kolmogorov turbulence. } 
\label{velocities}
\end{figure}

\subsection{X-ray spectroscopy}
Our temperature map in Fig.~\ref{maps} reveals a remarkably steep temperature gradient in the core of the cluster, with the azimuthally averaged temperature increasing from $\sim2$~keV at the center up to $\sim8$~keV at $r\sim20$~kpc. Outside $r\sim30$~kpc the temperature distribution is remarkably uniform with $kT\sim9.5$--10.5~keV. 

Zooming-in on the innermost $r=10$~kpc core of the cluster, using lower signal-to-noise (S/N=50) maps, the high temperature gradient in the core becomes even clearer (see Fig.~\ref{maps_HR}). Along the northwestern direction, the gradient of the projected temperature reaches d$kT/{\rm{d}}r\sim1$~keV/kpc. 
While we model the ICM using a single-temperature plasma model, the plasma in the innermost cluster core is multiphase. The soft emission of a multiphase plasma biases the best fit hydrogen column density, $N_{\rm H}$, low. The central panel of Fig.~\ref{maps_HR} shows low best fit $N_{\rm H}$ in the central four spatial bins, at radii $r\lesssim3.5$--4.5~kpc. Outside this innermost region, the best fit line-of-sight absorbing hydrogen column density distribution is relatively uniform and its value, $N_{\rm H}\sim4\times10^{21}$~cm$^{-2}$, is approximately two times higher than the column densities measured by the Leiden-Argentine-Bonn survey of Galactic \ion{H}{I} \citep{kalberla2005}. The difference is probably due to molecular gas and dust along our sight-line close to the Galactic plane. Fixing the $N_{\rm H}$ to $4\times10^{21}$~cm$^{-2}$ in our model (the best fit value determined in regions outside the cluster core), we fit a two-temperature model to the spectra extracted from the four central regions (regions where Fig.~\ref{maps_HR} shows a low $N_{\rm H}$). With respect to a fit with a single temperature plasma and free $N_{\rm H}$, for one additional free parameter, the $\chi^2$ of the fit improves by 210 for 4693 degrees of freedom. This best fit model indicates the presence of a $0.82\pm0.02$~keV and a $2.75\pm0.09$~keV plasma phase with a metallicity of $2.14\pm0.26$ Solar. Assuming the 0.82~keV phase is radiatively cooling, we also fit a model of a collisionaly ionized plasma and a cooling flow, cooling from 0.82~keV, and find a best fit mass deposition rate of $\dot{M}=0.97\pm0.12~M_{\odot}$~yr$^{-1}$. Similarly to other systems, such as M~87 \citep{werner2010,werner2013}, Perseus cluster \citep{sanders2007}, Centaurus cluster \citep{sanders2016}, the coolest gas phases are associated with H$\alpha$  filaments in the core of the cluster \citep{hamer2012}.

The metallicity map in the right panel of Fig.~\ref{maps} reveals a strongly centrally peaked iron abundance distribution. The apparent metallicity dip in the innermost cluster core (see right panel of Fig.~\ref{maps_HR}) disappears when fit with a two-temperature model, which indicates a relatively high central metallicity of $2.14\pm0.26$ Solar (see also the previous paragraph).  The metallicity distribution shows a relatively sharp discontinuity at the southeastern edge of the outer cold front at $r\sim43$~kpc and a trail of excess iron abundance extends to the north of the core. 

The projected entropy distribution is asymmetric (see the left panel of Fig.~\ref{thermo_maps}), with a clear discontinuity at the southern cold front and a low entropy `tail' extending northward. While, as indicated by our deprojection analysis (see the next paragraph), the true pressure peaks at the BCG, the projected pressure in the bright cluster core is biased low (see the right panel of Fig.~\ref{thermo_maps}). This bias is due to the underestimated density, which is the result of the adopted constant line-of-sight bin size\footnote{When looking at a spherically symmetric cluster with a centrally peaked density distribution, the effective line-of-sight length, from which the dominant fraction of photons is emitted, is smaller in the core and increases with radius.}. The detected azimuthal variations are, however, robust. Similar to the entropy distribution, the pressure distribution also shows a prominent excess to the north of the core.

Taking the X-ray surface brightness peak as the center of the system and assuming spherical symmetry, we determine the radial profiles of the deprojected spectral properties along five different azimuths, chosen based on the observed morphological characteristics (the azimuth angles are measured counterclockwise from the west): west (305--30 degrees), north (30--110), east (110--205), southeast (205--260), and south (260--305). Fig.~\ref{profiles1} shows the deprojected electron density ($n_{\rm e}$) and temperature ($kT_{\rm e}$) profiles and both the projected and deprojected iron abundance distributions (the deprojection analysis introduces a substantial noise into the measured Fe abundance values). Using these  measurements, we determine the total pressure ($P=nkT$) and entropy ($K=kT_{\rm e}/n^{2/3}_{\rm e}$) profiles shown in Fig.~\ref{profiles2}. In the innermost $r\sim$2~kpc region, the best fit iron abundance peaks at $\sim 2.3$~Solar and the entropy drops to $\sim2.6$~keV~cm$^2$ .

\begin{figure*}
\begin{minipage}{0.32\textwidth}
\hspace{-0.8cm}\includegraphics[width=1.05\textwidth,clip=t,angle=0.]{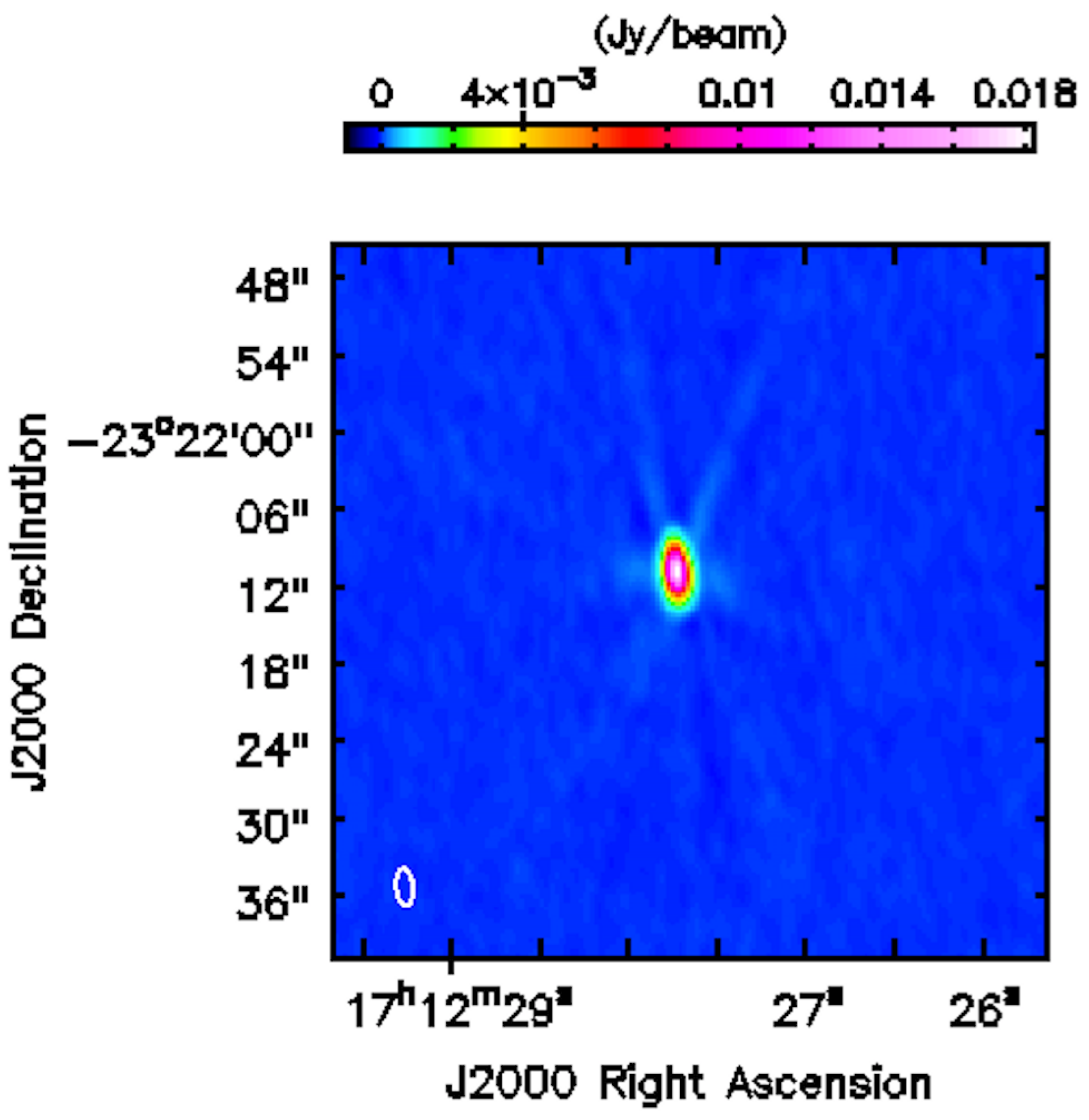}
\end{minipage}
\begin{minipage}{0.32\textwidth}
\hspace{-0.4cm}\includegraphics[width=1\textwidth,clip=t,angle=0.]{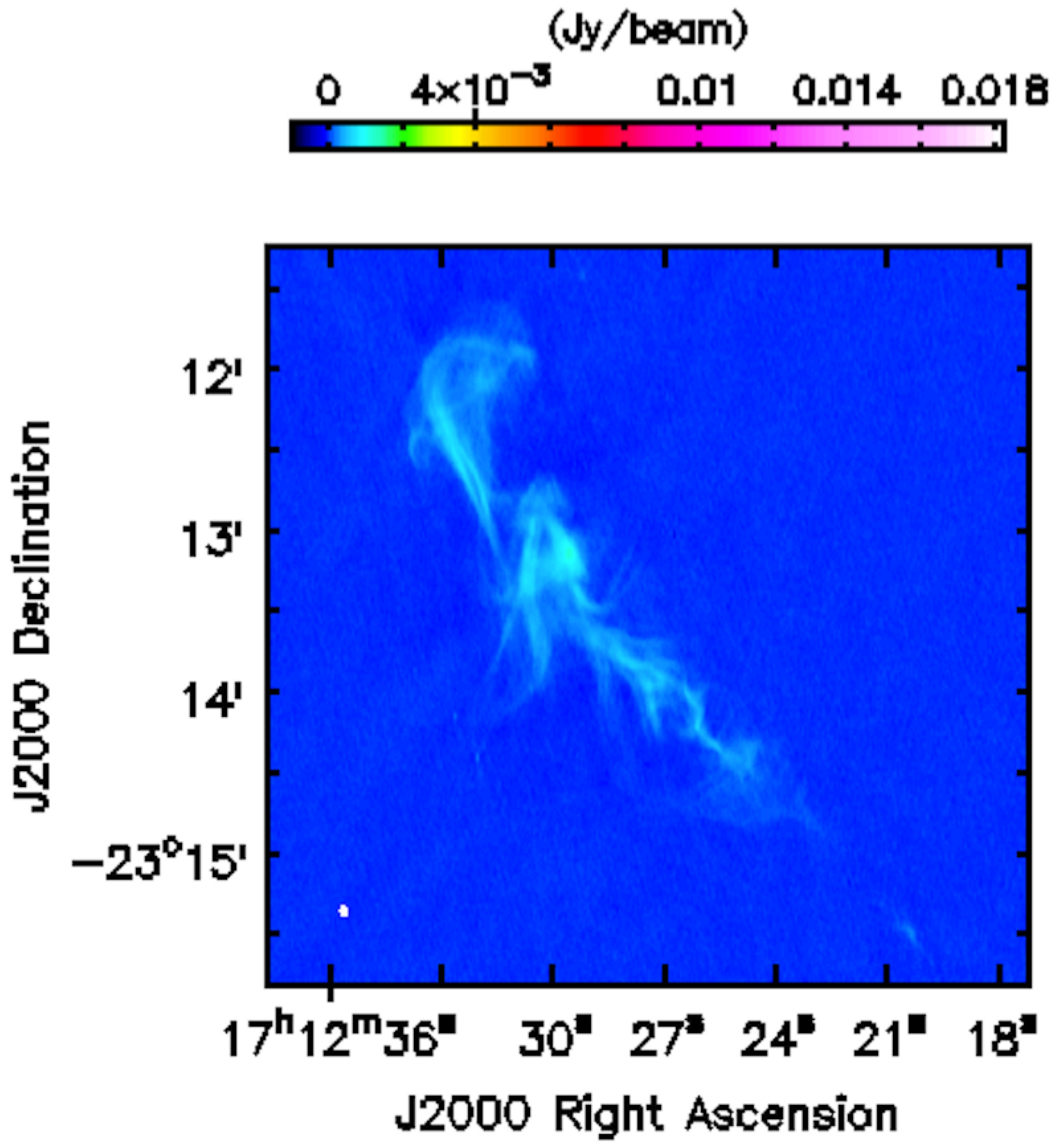}
\end{minipage}
\begin{minipage}{0.32\textwidth}
\includegraphics[width=1\textwidth,clip=t,angle=0.]{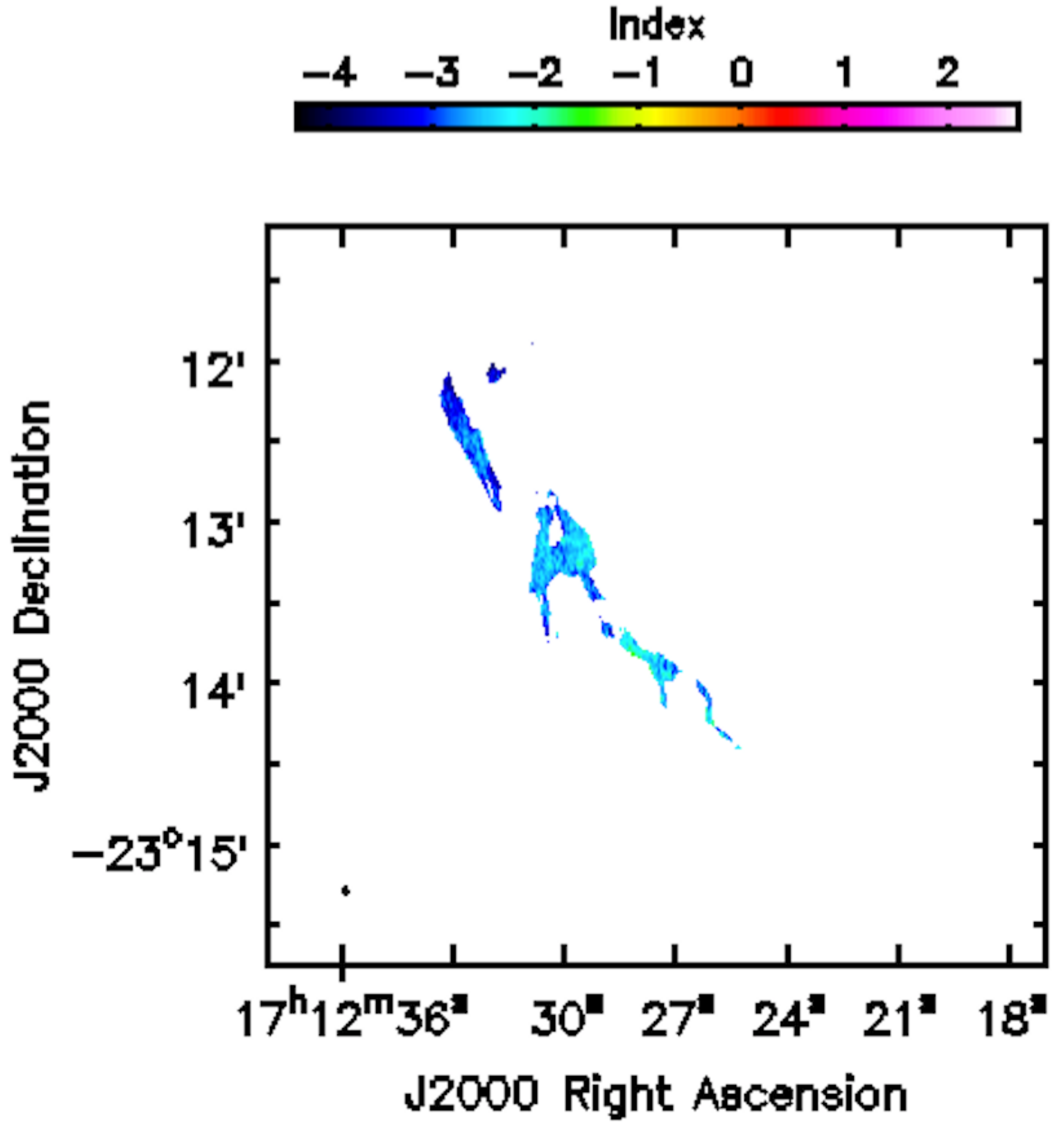}
\end{minipage}
\caption{{\it Left panel:} The BCG harbors a relatively faint, point-like, unresolved radio source with a 30 mJy flux density at 1.4~GHz. The white ellipse in the bottom left corner shows the beam size of $1.17\arcsec\times2.48\arcsec$. {\it Central panel:} A peculiar, filamentary radio source, located about 360~kpc to the north of the cluster core \citep[see also][]{murgia2010}. This object with no optical or X-ray counterpart is a likely radio phoenix, a source revived by adiabatic compression by gas motions in the ICM. {\it Right panel:} The spectral index map of the likely radio phoenix shows a relatively steep spectrum across the source.} 
\label{radioim}
\end{figure*}

\begin{figure}
\includegraphics[width=0.85\columnwidth,clip=t,angle=0.]{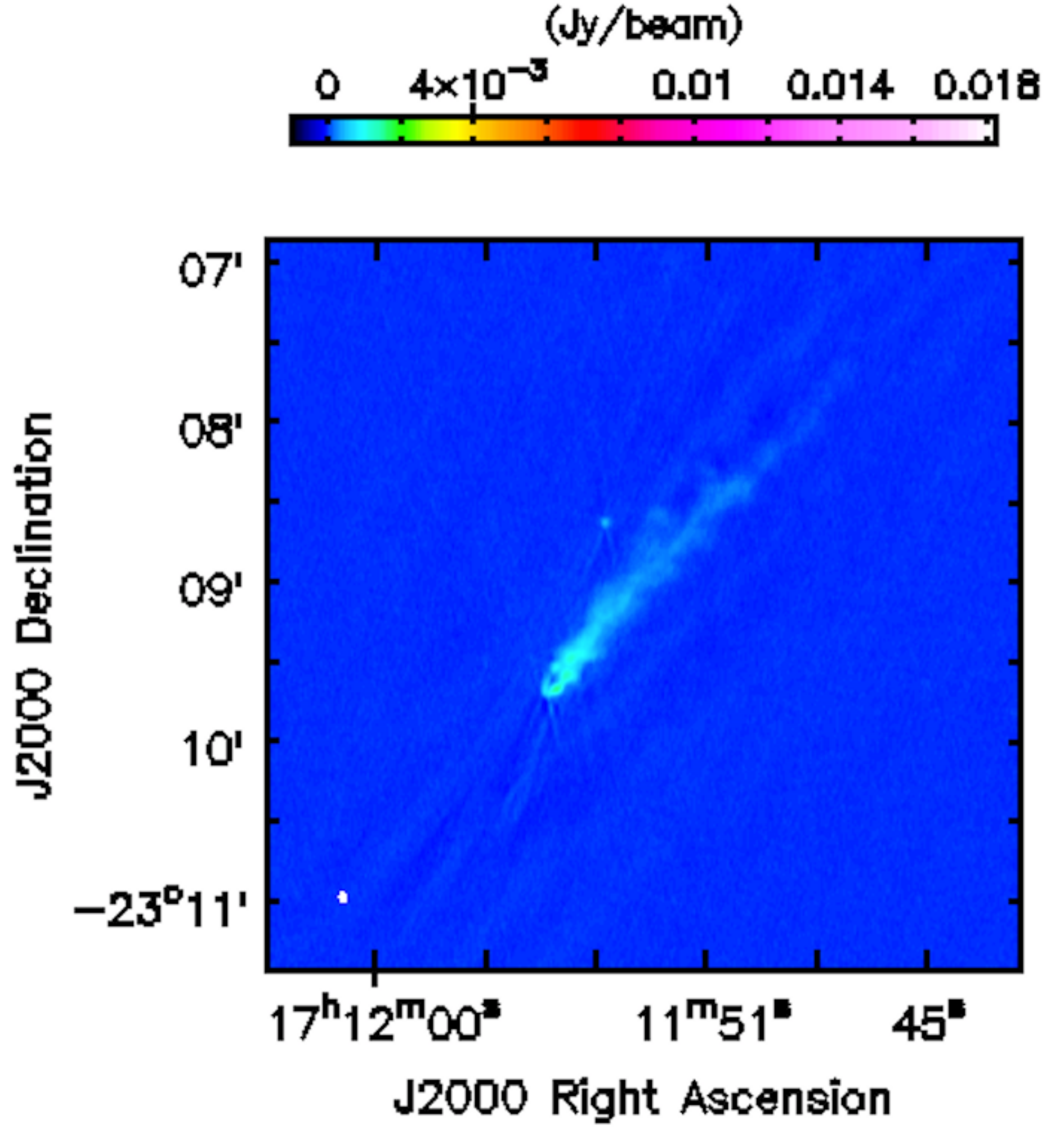}
\includegraphics[width=0.85\columnwidth,clip=t,angle=0.]{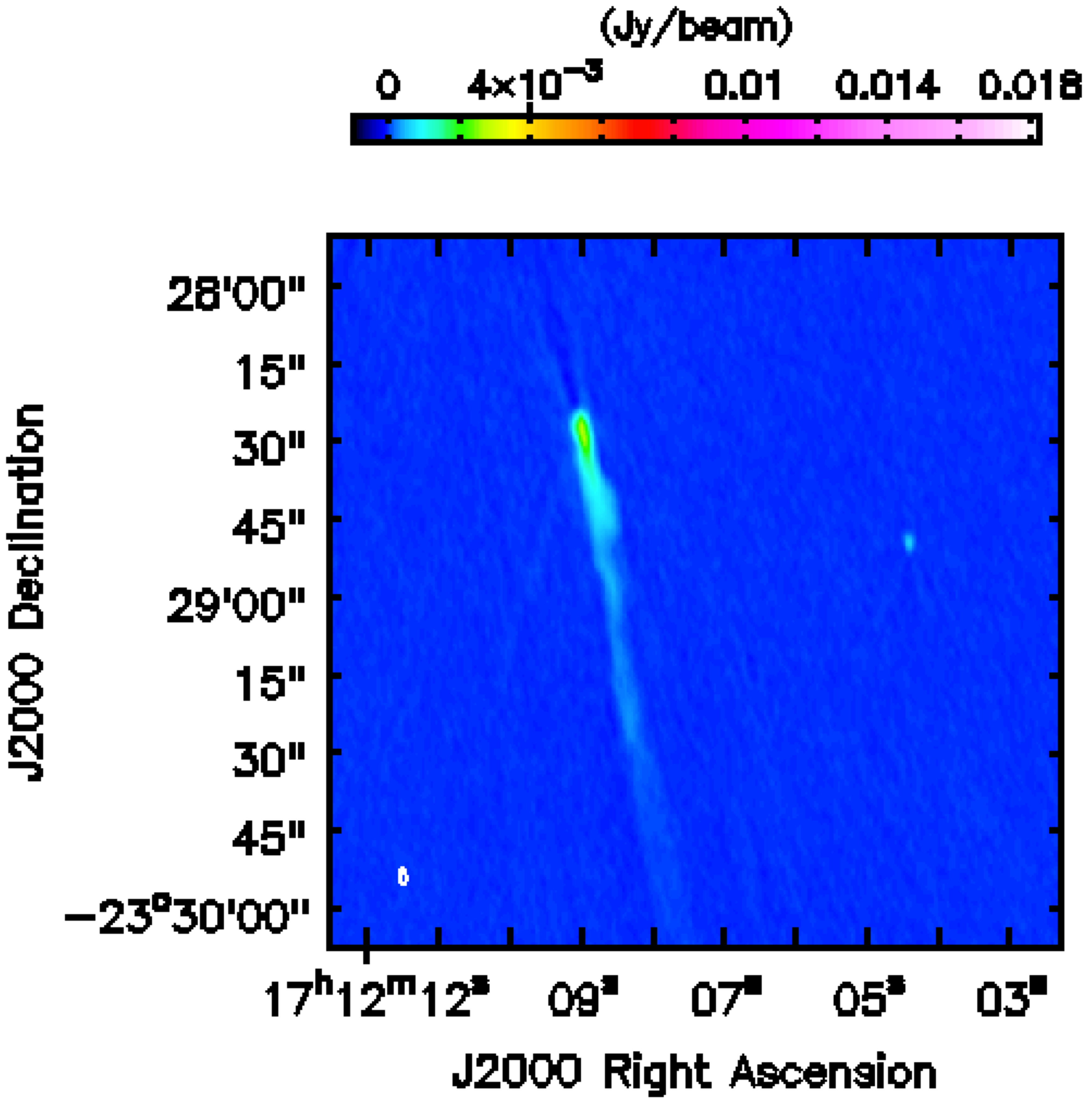}
\caption{The Ophiuchus cluster also harbors two bright narrow-angle tail radio galaxies \citep[see also][]{murgia2010}. The length of the observed radio emitting tails of these infalling galaxies is $\sim75$~kpc (upper panel) and $\sim40$~kpc (lower panel).  } 
\label{ntail}
\end{figure}

\subsection{The power-spectra of surface brightness fluctuations}
\label{powerspectra}

Following the methods described in \citet{churazov2012}, \citet{zhuravleva2015,zhuravleva2016}, and \citet{arevalo2015}, we perform an analysis of the power-spectra of the surface brightness fluctuations in the Ophiuchus cluster. Using theoretical arguments supported by numerical simulations, \citet{zhuravleva2014L} and \citet{gaspari2014} show that in galaxy clusters, where the gas motions are subsonic, the root mean squared amplitudes of the density and line-of-sight-component velocity fluctuations are proportional to each other on each scale k$^{-1}$, $\delta \rho_{\rm{k}} /\rho_{0} \approx \eta V_{\rm{1D,k}} / c_{\rm{s}}$, where $\rho_{0}$ is the unperturbed gas density, $c_{\rm{s}}$ is the sound speed and $\eta\approx1$ is a proportionality coefficient. Cosmological simulations show that $\eta\approx1$ with a scatter of 30 per cent \citep{zhuravleva2014L} and hydrodynamic simulations confirm that the proportionality coefficient is of the order of unity as long as conduction is completely suppressed \citep{gaspari2014}. 

To avoid complications due to the steep temperature gradient in the core of the cluster, we analyzed an image in the annular region of $r=1.5$--5~arcmin around the surface brightness peak, where the ICM appears approximately isothermal (see the left panel of Fig.~\ref{maps}). The image was extracted in the 1.5--3.5~keV energy range, where the emissivity is approximately independent of temperature and proportional to the density squared.  We ignored the soft band where we are affected by the high Galactic column density. To remove the global surface brightness gradient, we divided the cluster image by a model of an ``unperturbed'' surface brightness distribution (see Fig.~\ref{modelsub}). Our simplest assumption is a spherically symmetric $\beta$-model which, however, does not account for the prominent global asymmetry. To remove the asymmetry, as shown in Fig.~\ref{modelsub}, we experimented with $\beta$-models patched on several different scales \citep[see][for details]{zhuravleva2015}. 

Assuming volume filling turbulence and a sound speed of 1500~km~s$^{-1}$, we calculate the velocity power spectrum from the spectrum of the density fluctuations obtained by using a range of models for the underlying global density distribution. We find that after accounting for the azimuthal variation in the underlying model, the derived volume filling velocities are $V_{\rm{1D,k}}\lesssim100$~km~s$^{-1}$ on scales $\lesssim100$~kpc, see Fig.~\ref{velocities}.  

To determine the nature of the southern surface brightness excess (see Fig.~\ref{large_IM}), we used a method introduced in \citet{arevalo2015} and \citet{zhuravleva2016}, and analyzed the statistical properties of the emissivity fluctuations in two energy bands: 1.5--3.5~keV and 3.5--7.5~keV. For the gas temperatures in the Ophiuchus cluster, we would expect that while `isobaric' gas perturbations, associated with slow displacement of gas, will result in a ratio of the power-spectra derived in the hard and soft energy bands of $R\sim0.9$, `adiabatic' perturbations, associated with weak shocks, will result in $R\sim1.1$. Our measured ratios, after accounting for the large statistical uncertainties, are $R>1$ consistent with `adiabatic' perturbations, suggesting the possible presence of a weak shock or of a gravitational potential well of a separate sub-halo moving through the cluster. The statistics associated with this feature is, however, low and a more direct confirmation of the nature of this feature will require deeper observations. 

\subsection{Radio properties}
\label{radio}

We detect a point-like radio source associated with the center of the BCG with 
a flux density of 30 mJy and spectral index $\alpha=-0.55$. 
The source is both compact and optically thin, typical of very low
level accretion. As can be seen from the left panel of Fig.~\ref{radioim}, we find no evidence of
extended structure near the BCG above the rms level of 10 $\mu$Jy. The upper limit on the polarization of the source is 0.2 per cent.

Our radio data also show two head-tail radio galaxies (see Fig.~\ref{ntail}) previously studied by \citet{murgia2010} and \citet{govoni2010}. However, the brightest (630~mJy) and perhaps most interesting radio source in the Ophiuchus cluster is a peculiar, 200 arcsec long (120~kpc at the redshift of the cluster), filamentary object (see the central panel of Fig.~\ref{radioim}), located about 360~kpc to the north of the cluster core.  The object has been previously identified in GMRT data by \citet{murgia2010} as a possible unusual radio relic. We find a steep spectral index of $\alpha \sim -1.8$ to $-2.5$ that is relatively uniform across the source (see the right panel of Fig.~\ref{radioim}). This index is steeper than the slope of $\alpha=-1.01\pm0.03$ measured between 74~MHz and 1.4~GHz by \citet{murgia2010}, indicating the presence of a break at $\sim1$~GHz. The source shows signs of polarization, with the brightest knot, located near the middle of the source, being polarized to roughly 18 per cent. The object has no optical/near-infrared counterpart. Because it is offset by about 10 arcmin from the aim-point of the {\it Chandra} observations, the object has not been exposed by the ASIC-I array, however, it has been covered by the CCD-6 (ACIS-S2) during part of our observation. These CCD-6 data and an archival short {\it XMM-Newton} observation do not reveal an X-ray counterpart for the radio source. The quality of the X-ray data does not allow us to investigate the thermodynamic properties of the ICM at the location of the object in detail.   

The core of the Ophiuchus cluster also harbors a faint radio minihalo \citep[see][and our discussion below]{govoni2009,murgia2009}, however, our high resolution JVLA observation in the A configuration is not sensitive to the faint extended emission.

\section{Discussion}
\label{discussion}

\subsection{Dynamical activity and the truncated cooling core}

The $\sim38$~kpc offset of the centroid of the large scale X-ray emission distribution from the central BCG and the prominent southern surface brightness excess indicate that the gravitational potential in the cluster core has been strongly perturbed. The maps of projected thermodynamic properties support this picture. In a galaxy cluster in approximate hydrostatic equilibrium, with subsonic gas motions, the pressure distribution traces the gravitational potential of the system. However, contrary to most cooling core clusters, the pressure distribution in the Ophiuchus cluster is highly azimuthally asymmetric around the BCG and displays a prominent region of excess pressure in the north. These results strongly indicate that the BCG, which presumably traces the dark matter density peak, is offset from the cluster's global center of mass.  Such a perturbed gravitational potential is most likely the result of a relatively recent encounter with an in-falling sub-cluster.  

The multiple concentric cold fronts in the core of the cluster are also most likely due to gas sloshing in a changing gravitational potential \citep[e.g.][]{ascasibar2006,markevitch2007} which, following an encounter with a sub-cluster, is expected to be evolving towards a global equilibrium. While the southern outer cold front, at $r\sim43$~kpc from the BCG, indicates southward gas motions, the prominent eastern inner cold front, at $r\sim5$~kpc from the BCG, shows that the gas in the cluster core is swinging northeastward, most likely following the innermost part of the dark matter potential. As the dense low entropy ICM in the cluster core moves through the ambient ICM, it encounters ram-pressure and, as proposed by \citet{million2010}, the outer layers of the cool core get stripped, resulting in the highly unusual steep temperature and metallicity gradients \citep[for comparison see the entropy profiles in][]{cavagnolo2009,panagoulia2014}. The entropy and metallicity distributions in Fig.~\ref{maps}--\ref{thermo_maps} also show clear evidence for such ram-pressure stripping. On the large scale maps, we see an excess of low-entropy, high-metallicity gas northwest of the core of the cluster. On smaller scales, we see a similar excess along a perpendicular axis, to the southwest. In the innermost cluster core (see Fig.~\ref{maps_HR}), we see an excess of cooler gas to the southeast. 
The observed disruption of the cool core in the Ophiuchus cluster is particularly interesting in the light of simulations, which indicate that cool cores are remarkably resilient and survive late major mergers \citep{burns2008,poole2008}. 

The southern surface brightness excess, seen at the bottom of the residual image in the right panel of Fig.~\ref{large_IM}, is an obvious candidate for a sub-cluster that could have encountered the cluster core relatively recently. The appearance of its southern edge is suggestive of a bow-shock like shape, which would indicate a merger taking place close to the plane of the sky. Furthermore, as discussed in Section~\ref{powerspectra}, the cross spectrum analysis also suggests the possible presence of a weak shock.   
Assuming a merger in the plane of the sky, occurring at the sound speed in the $kT=8.5$~keV ICM ($c_{\rm s}=1500$~km~s$^{-1}$), we estimate that the closest passage to the cluster core would have occurred less than  $\approx200$~Myr ago. 
A recent study of spectroscopic redshifts by \citet{durret2015} indicates that the $1.1\times10^{15}~M_{\odot}$ Ophiuchus cluster is not undergoing a major merger and the BCG is close to the center of the global gravitational potential, where the BCG velocity is consistent with the mean cluster velocity $\Delta v =47\pm97$~km~s$^{-1}$. However, a merger with a $M\sim10^{14}~M_{\odot}$ system would still be consistent with the optical data.  

Although our data clearly show that the Ophiuchus cluster is dynamically active, it is unlikely that the truncation of the cool core has been caused by a single, recent merging event. Comparison to numerical simulations \citep[e.g.][]{ascasibar2006,zuhone2011,rodiger2011} indicates that the observed complex, developed sloshing patterns are likely to have been triggered by an encounter over 1 Gyr ago. The merger with the southern sub-cluster might therefore be the latest in a series of systems merging with the Ophiuchus cluster.

\subsection{The radio phoenix}
\label{phoenix}
Another indication of dynamical activity in the Ophiuchus cluster are the two narrow-angle tail radio sources \citep[see Fig.~\ref{ntail} and][]{murgia2010} which are infalling towards the cluster center, and the peculiar, filamentary radio source detected at $r\sim360$~kpc (10~arcmin) north of the BCG (see the central panel of Fig.~\ref{radioim}). This relic-like source has no optical/near-infrared or X-ray counterpart. 
The object lacks a clear radio core, and its spectral index of $\alpha\sim-2$ is similar to that of radio relics \citep[see also][]{murgia2010}. However, the relatively small size (about $120\times30$~kpc), unusual morphology, high surface brightness, and the central location within the Ophiuchus cluster make it unlikely that this source is a radio relic. It appears more similar to {\it radio phoenixes} \citep{kempner2004,vanweeren2009,degasperin2015}, sources revived by adiabatic compression by a shock or gas motions propagating through the ICM as shown by \citet{ensslin2001} and \citet{ensslin2002}.  The vortex-like structure at its northern tip might be associated with a $L\sim15$~kpc turbulent eddy in the ambient ICM. Contrary to the recently identified radio phoenix in Abell 1033 \citep{degasperin2015}, the source in the Ophiuchus cluster shows signs of polarization, which may be due to its larger distance from the cluster center resulting in a smaller amount of intervening matter. 

\subsection{The cold fronts}
\label{coldfronts}

The inner eastern cold front (Fig.~\ref{s_IM}) is remarkably sharp, with a width smaller than 1.5 kpc. This is the tightest measured upper limit on a physical width of a cluster cold front. However, because the ICM density in the cluster core is high, the Coulomb mean free path for particles diffusing from the cooler bright, dense side of the discontinuity to the hotter outer ICM is about an order of magnitude smaller than this upper limit. Diffusion has previously been shown to be suppressed at cold front interfaces with upper limits in Abell 3667 and the Virgo cluster smaller or comparable to the Coulomb mean free path \citep{vikhlinin2001b,markevitch2007,werner2016}. 

The `tongue'-like extension to the north (Fig.~\ref{s_IM}) may be due to the onset of a Rayleigh-Taylor instability, which develops when the displaced, dense gas subjected to ram-pressure from the ambient ICM spreads sideways, eventually sprouting a tongue of low entropy material which separates and starts sinking towards the minimum of the gravitational potential. Note the remarkable morphological similarity of this feature (although on a much smaller physical scale) to the results of the simulation in Fig.~7 of \citet[][]{ascasibar2006}. 

The outer southern cold front appears significantly more disturbed, with multiple edges and azimuthal variation. Fig.~\ref{m_IM} shows that while the southern edge of the front is relatively sharp (width smaller than 5.3~kpc), its southeastern side appears smeared. Azimuthal variations and smearing have also been observed at the prominent cold fronts in the Virgo and Centaurus clusters by \citet{werner2016} and \citet{sanders2016}. They propose that the smearing is most likely due to Kelvin-Helmholtz instabilities, which have previously also been observed in NGC~7618 and UGC~12491 \citep{rodiger2012}.   Constrained hydrodynamic simulations reproducing the Virgo cluster by \citet{rodiger2013} demonstrate that viscosities of about 10 per cent of the Spitzer value or larger are very efficient at preventing the development of Kelvin-Helmholtz instabilities, indicating that the viscosity is suppressed in that system. Magnetohydrodynamic simulations by \citet{zuhone2011,zuhone2013} show that magnetic fields provide only partial protection against hydrodynamic instabilities. The apparent breakup of the southern cold front is thus consistent with the previous results which suggest that the viscosity of the ICM is suppressed with respect to the temperature dependent Spitzer value. 

While the multiple surface brightness edges to the south of the cluster center, seen in the right panel of Fig.~\ref{m_IM}, could be due to Kelvin-Helmholtz instabilities, as discussed by \citet{rodiger2013}, they could also be due to gas depletion by amplified magnetic fields underneath the cold front, as discussed in \citet{werner2016}. Simulations of \citet{zuhone2011,zuhone2015} show that gas sloshing may result in wide bands of amplified magnetic fields extending relatively far below cold fronts. The increased magnetic pressure in such layers may push gas out, decreasing its density, creating surface brightness edges associated with the magnetized bands.

\subsection{Turbulence and thermal conduction in the hot ICM}
\label{turbulence}

Even though our observations clearly show that the core of the Ophiuchus cluster is strongly dynamically active, the amplitude of density fluctuations in the $r=1.5$--5~arcmin (53--177~kpc) region is relatively low, indicating velocities smaller than $\sim100$ km~s$^{-1}$ on scales $\lesssim100$~kpc. Given that the cooling core is offset from the global center of mass and the sub-cluster that initiated the unusually strong sloshing is likely to have passed close to the cluster center, we would expect the gas motions associated with the dynamical activity to be higher and to have cascaded down to a volume filling turbulence in the investigated region. Moreover, such low ICM velocities would be surprising in a disturbed cluster with a galaxy velocity dispersion of $\sim954\pm58$ km~s$^{-1}$ \citep{durret2015}. Viscosity is unlikely to suppress gas motions on the relatively large scales that we are probing \citep[see e.g.][]{gaspari2014}, and the morphology of cluster cold fronts indicates that the viscosity in the ICM is likely lower than the temperature dependent Spitzer value by more than an order of magnitude \citep[see Sect.~\ref{coldfronts} and][]{rodiger2013,rodiger2015,werner2016}. Our relation between the power spectrum of the density fluctuations and the power-spectrum of velocity fluctuations was derived under the assumption that thermal conduction in the ICM is fully suppressed. However, thermal conduction could damp the density fluctuations by a factor $\sim2$--3 on the observed scales, while leaving the velocity cascade unaltered \citep{gaspari2014}. Such damping would then result in our underestimate of velocities. 

The large temperature gradient, of the order of d$kT/{\rm{d}}r\sim1$~keV/kpc, in the core of the Ophiuchus cluster indicates that the thermal conduction is suppressed along the radial direction. Such suppression might be the result of azimuthal magnetic fields that will, in the absence of significant turbulence, arise in cooling cores due to heat-flux driven buoyancy instability \citep[HBI;][]{quataert2008,parrish2008,parrish2009,parrish2010}. Gas sloshing can also result in azimuthally aligned magnetic fields \citep{zuhone2011}. Observations of cold fronts \citep{ettori2000,xiang2007} and merging clusters of galaxies \citep{markevitch2003} also indicate that thermal conduction in the ICM is significantly suppressed (by a factor of order $\sim10^2$) with respect to the Spitzer value.
The alignment of magnetic field lines alone, however, would not explain the observed suppression of the heat flux. Comparisons of numerical simulations that include anisotropic conduction to observations indicate that thermal conduction is also suppressed by over an order of magnitude {\it along} the magnetic filed lines \citep{zuhone2013,werner2016}. However, because thermal conduction is a strong function of temperature ($\propto T^{5/2}$), even a suppressed conductivity would be relatively efficient in the $\sim9$~keV ICM outside of the cooling core of the Ophiuchus cluster. Conduction could, in principle, contribute both to the observed isothermality and to the damping of density fluctuations. It would suppress the perturbations preferentially on smaller scales, steepening the observed power-spectrum. Although the observed velocity power-spectrum in Fig.~\ref{velocities} appears steeper than the Kolmogorov spectrum, the large systematic uncertainties due the subtraction of the underlying model do not allow us to make a firm conclusion about the slope.

A relatively low level of surface brightness fluctuations is also measured  in the $kT\sim3.5$~keV AWM~7 cluster, where conduction is expected to be much less significant \citep{sanders2012}. It is, however, possible that this system is dynamically more relaxed and does therefore not harbor strong volume filling turbulence. Conduction would probably also have a less significant effect on the surface brightness fluctuations in the cool cores of the Perseus and Virgo clusters  \citep[see][]{zhuravleva2014,zhuravleva2015,arevalo2015}.  
Comparison of the power-spectra of surface brightness fluctuations and direct line broadening measurements with calorimeter type high-resolution X-ray spectrometers will provide important quantitative constraints on transport processes and the microphysics of the ICM \citep[see the discussion in][]{gaspari2014,zuhone2015}.

\subsection{The concave surface brightness discontinuity: giant AGN outburst or merger related gas dynamics? }

Concave surface brightness discontinuities, similar to the one observed at $r\sim120$~kpc southeast of the core of the Ophiuchus cluster, but on smaller physical scales, are also observed in the Perseus cluster, Abell~1795, Abell 2390, and Centaurus cluster, where they were interpreted as the inner walls of cavities resulting from AGN activity \citep{walker2014,sanders2016}. Low frequency radio observations of the Ophiuchus cluster performed by the GMRT also reveal steep spectrum radio emission beyond this feature \citep[see the radio source E in][]{murgia2010}. The radius of curvature of the observed concave feature is $\sim 180$~kpc, therefore, if it is associated with the inner wall of a spherical cavity, then the $pV$ work required to displace the ICM is about $5\times10^{61}$~ergs \citep[comparable with the most powerful AGN outburst known in MS~0735.6+7421;][]{mcnamara2005,vantyghem2014}. However, a buoyantly rising AGN inflated cavity would more likely have an ellipsoidal shape and a smaller size, requiring less but still a substantial amount of AGN power. Such a powerful AGN outburst would heat and displace a large amount of ICM in the cluster core and, in combination with the ongoing merger, contribute sinificantly to the disruption of the cooling core \citep[see e.g.][]{ehlert2011}. 

However, the extremely steep metallicity and entropy gradients in the Ophiuchus cluster would most likely get erased by such a powerful AGN outburst, arguing against this scenario. It is more likely that the discontinuity is the result of merger related gas dynamics. Similar concave discontinuities appear at certain stages of mergers in simulations of infall of gas-rich subclusters into cooling core clusters \citep[see Fig. 22 in][]{ascasibar2006}.

\subsection{Suppressed AGN feedback}

Because of the ram pressure experienced by the ICM (in contrast to the stars and dark matter associated with the BCG, which are effectively collisionless), the gas motions in the changing gravitational potential well can lead to the separation of the ICM density peak from the BCG. Such a separation is clearly observed in the Ophiuchus cluster, where the X-ray surface brightness peak is displaced from the center of the BCG by 2.2 kpc in projection and spatially coincides with H$\alpha$ nebulae. This dense displaced gas displays multi-temperature structure consistent with a radiative cooling rate of $\dot{M}=0.97\pm0.12~M_{\odot}$. Because this gas is displaced from the vicinity of the AGN, it can continue cooling without triggering an AGN feedback response. Similar cooling, offset from the AGN, has been seen in Sersic 159-03, Abell~3444, Abell~1991, and Abell~2146 \citep[e.g.][]{werner2011,hamer2012,mcdonald2015,russell2012,canning2012}. The lack of observed X-ray cavities in the cluster core and the weak point-like radio emission, lacking lobes or jets, indicate that the AGN may currently be largely dormant, which is consistent with the cooling taking place offset from the central supermassive black hole. The innermost surface brightness discontinuity, just 2~kpc west of the X-ray peak, separates the cooling multiphase gas from the ambient ICM.

\citet{mazzotta2008} propose that turbulence driven by gas sloshing reaccelerates preexisting cosmic rays from the central AGN producing diffuse radio mini-haloes. The presence of the radio mini-halo in the Ophiuchus cluster, discovered by \citet{govoni2009} using VLA data at 1.4 GHz in the D configuration, is consistent with this picture. \citet{murgia2009} concluded that the emissivity of the mini-halo of the Ophiuchus cluster (along with the mini-haloes in Abell 1835 and Abell 2029) is low, similar to the emissivities of radio haloes of merging clusters rather than to other previously known mini-haloes. Within the framework of the turbulent re-acceleration scenario, the relatively low surface brightness of the mini-halo in such a dynamically active cluster core implies the lack of a substantial amount of preexisting cosmic rays, consistent with a picture of a largely dormant AGN.

\section{Conclusions}
\label{conclusions}

\begin{itemize}

\item The Ophiuchus cluster hosts a truncated cool core, with a temperature increasing from $kT\sim1$~keV in the center to $kT\sim9$~keV at $r\sim30$~kpc. Beyond $r\sim30$~kpc the ICM appears remarkably isothermal. 

\item The core is dynamically disturbed with multiple sloshing induced cold fronts, with indications for both Rayleigh-Taylor and Kelvin-Helmholtz instabilities. 
The sloshing is the result of the strongly perturbed gravitational potential in the cluster core, with the BCG displaced southward from the global center of mass. 

\item The residual image reveals a likely sub-cluster south of the core at the projected distance of $r\sim280$~kpc. 

\item The cluster harbors a peculiar, filamentary radio source with no optical or X-ray counterpart, located about 360~kpc to the north of the cluster core. The object is a likely radio phoenix, a source revived by adiabatic compression by gas motions in the ICM. 

\item Even though the Ophiuchus cluster is strongly dynamically active, the amplitude of density fluctuations outside of the cooling core is low, indicating velocities smaller than $\sim100$ km~s$^{-1}$. The density fluctuations might be damped by thermal conduction in the hot and remarkably isothermal ICM, resulting in our underestimate of gas velocities. 

\item  We find a surprising, sharp surface brightness discontinuity that is curved away from the core, at $r\sim120$~kpc to the southeast of the cluster center. We conclude that this feature is most likely due to gas dynamics associated with a merger and not a result of an extraordinary AGN outburst. 

\item The cooling core lacks any observable X-ray cavities and the AGN only displays weak, point-like radio emission, lacking lobes or jets, indicating that currently it may be largely dormant. The lack of strong AGN activity may be due to bulk of the cooling taking place offset from the central supermassive black hole. The observed unchecked cooling where a largely dormant AGN is offset from the cooling flow solidifies the idea that AGN play a key role in maintaining the cooling/heating balance in cluster cores.

\end{itemize}

\section*{Acknowledgments}
Support for this work was provided by the National Aeronautics and Space Administration through Chandra Award Number GO4-15126X and GO4-15121X issued by the Chandra X-ray Observatory Center, which is operated by the Smithsonian Astrophysical Observatory for and on behalf of the National Aeronautics Space Administration under contract NAS8-03060. This work was supported in part by the US Department of Energy under contract number DE-AC02-76SF00515.
The authors thank Georgiana Ogrean and John ZuHone for discussions, and Evan Million and Steven Ehlert for their help with the observing proposal.

\bibliographystyle{mnras}
\bibliography{clusters}

\end{document}